\documentclass[aps,prd,amsfonts,amssymb,amsmath,superscriptaddress,nofootinbib]{revtex4-1}

\usepackage{ulem,color,wrapfig,graphicx,bm,float,braket}

\begin{document}
 
\title{$F(R)$ gravity in the early Universe:\\ Electroweak phase transition and chameleon mechanism}

\author{Taishi Katsuragawa}
\email{taishi@mail.ccnu.edu.cn}
\affiliation{Institute of Astrophysics, Central China Normal University, Wuhan 430079, China}
\author{Shinya Matsuzaki}
\email{synya@jlu.edu.cn}
\affiliation{Center for Theoretical Physics and College of Physics, Jilin University, Changchun, 130012, China}
\author{Eibun Senaha}
\email{senaha@ibs.re.kr}
\affiliation{Center for Theoretical Physics of the Universe, Institute for Basic Science (IBS), Daejeon 34051, Korea}

\begin{abstract}
It is widely believed that the screening mechanism is an essential feature for the modified gravity theory.
Although this mechanism has been examined thoroughly in the past decade,
their analyses are based on a conventional fluid prescription for the matter-sector configuration.
In this paper,
we demonstrate a new formulation of the chameleon mechanism in $F(R)$ gravity theory,
to shed light on quantum-field theoretical effects on the chameleon mechanism as well as the related scalaron physics, 
induced by the matter sector.
We show a possibility that the chameleon mechanism is absent in the early Universe based on a scale-invariant-extended scenario beyond the standard model of particle physics, 
in which a realistic electroweak phase transition, yielding the right amount of baryon asymmetry of Universe today, simultaneously breaks the scale invariance in the early Universe. 
We also briefly discuss the oscillation of the scalaron field and indirect generation of non-tensorial gravitational waves 
induced by the electroweak phase transition.
\end{abstract}

\maketitle



\section{Introduction}

Dark energy problem for the accelerated expansion of the Universe is one of the biggest mysteries in modern physics. 
Despite observational successes of the $\Lambda$CDM model in the framework of the general relativity, 
this model suffers from the cosmological constant problem,
so the cosmological constant is still a phenomenological parameter.
The modified gravity theories can provide us with alternative explanations for the dark energy, 
instead of ad hoc addition of the constant term.
From the theoretical and phenomenological viewpoints, 
a variety of modified gravity theories has been proposed so far (for example, see \cite{Clifton:2011jh,Nojiri:2010wj}),
in which the new dynamics in the gravity sector is responsible for the origin of the late-time cosmological acceleration.
Moreover, the challenges to test the beyond-standard in gravitational physics have received much attention
\cite{Joyce:2014kja,Berti:2015itd,Koyama:2015vza,Bull:2015stt}.

In order for the modified gravity theory to be realized in nature,
we have to pay attentions to the phenomenon at the smaller scale.
Modifications for the cosmological scale also affect the predictions for galaxy clusters, galaxies, and the solar system,
and they lead inconsistent results with the observations.
Thus, we require the screening mechanism \cite{Brax:2013ida,Burrage:2017qrf} to restore the general relativity at certain scales,
which suggests that the recovery of the general relativity must show the environment dependence.

The chameleon mechanism \cite{Khoury:2003aq,Khoury:2003rn} is one of the screening mechanisms,
and it appears in the scalar-tensor theory and $F(R)$ gravity theory that include an extra scalar field
which we call scalaron.
The potential term of the scalaron field includes the coupling to the trace of energy-momentum tensor $T^{\mu}_{\ \mu}$
which comprises the matter fields other than the scalar field itself.
Trace of the energy-momentum tensor controls the mass of the scalaron field, 
which is very large in a high-density region at the local scale and very small in a low-density region at the cosmological scale.
If the mass of the scalar field is large enough, the propagation is suppressed,
and the modified gravity theories restore the general relativity.

It should be noted that
the chameleon mechanism does not work if the trace of energy-momentum tensor vanishes.
In the previous research by two of authors \cite{Katsuragawa:2017wge},
the chameleon mechanism in the $F(R)$ gravity with the energy-momentum tensor which consists of the standard-model-particles was examined. 
Based on the conventional fluid approach
where one can approximately compute the energy density and pressure as free particles in the grand canonical ensemble,
we constructed the energy-momentum tensor.
We found that the trace of energy-momentum tensor is proportional to the mass-squared and the temperature-squared, 
$T^{\mu}_{\ \mu} \propto m^{2} T^{2}$,
and that the chameleon mechanism remains to work in the high-temperature epoch
if the particles have even a tiny mass
though they behave approximately as radiation.

Towards the complete understanding for the cosmic history of the scalaron field, it is mandatory to take into account the thermal history of the matter sector.
Thus, the formulation of the chameleon mechanism in the quantum field theory has a potential significance for the cosmology,
which may give us a powerful tool to study the modified gravity in the early Universe.
We also expect to find the unique phenomena related to the chameleon mechanism, 
which allows us to search the new physics originated from the modified gravity.

Based on the previous result \cite{Katsuragawa:2017wge},
the trace of energy-momentum tensor monotonically increases as it goes back to the past in the early Universe.
Thus, the chameleon mechanism has a larger effect on the mass of the scalaron field in the earlier Universe.
However, there is a crucial room to discuss the weakened or disabled chameleon mechanism in the early Universe, which was not evaluated in the previous work: 
to our common knowledge in particle physics and cosmology, 
the electroweak phase transition (EWPT) is supposed to have taken place 
when the temperature dropped to the EW scale of $\mathcal{O}(100 [\mathrm{GeV}])$.
If we believe the scale invariance at the classical level of a quantum field theory which we employ, 
and if it is to be broken explicitly and spontaneously by the EWPT 
through the dimensional transmutation and inclusion of thermal loop effects,  
the potential of the scalaron field receives a sudden and gigantic effect through the chameleon mechanism.
If we have a scale invariance, which implies $T^{\mu}_{\ \mu}$ vanishes, before the EWPT,
$T^{\mu}_{\ \mu}$ changes from zero to nonzero at the moment of the EWPT.
Hence, the scalaron would dramatically acquire a large mass after the EWPT.

On the other hand, as was argued in~\cite{Blas:2011ac} and \cite{Ferreira:2016wem,Ferreira:2016kxi}, 
one might think that the chameleon mechanism may not significantly be affected 
even at the quantum loop level
as long as the scale-symmetry breaking happens to be only spontaneous, 
where an ad hoc explicit-scale breaking arising from the renormalization procedure could be gone 
by a scale-invariant regularization method~\cite{Ghilencea:2015mza,Ghilencea:2016ckm}, 
hence $T^{\mu}_{\ \mu} =0$ even after the EWPT. 
However, inclusion of finite-temperature effects, 
which has not been addressed in the above literature, 
would necessarily cause an explicit-scale breaking for the target-matter sector
because the temperature is the physical scale, in contrast to the renormalization scale. 
Thereby, we would anticipate 
that the EWPT in the early hot Universe along classically scale-invariant scenarios 
surely yields nonzero $T^{\mu}_{\ \mu}$, (at least) by the explicit-scale breaking due to the presence of the thermal bath, 
which keeps nonzero even after the EWPT.

As a first step for such a completion of the chameleon mechanism in the early Universe, 
in this paper, 
we discuss the scalaron dynamics coupled to a class of scale-invariant two-Higgs doublet model (SI-2HDM),  
chosen as a referenced realistic scenario in terms of thermal history in the early Universe. 
It has been shown in the SI-2HDM~\cite{Fuyuto:2015vna} 
that the thermal effect arising from the presence of heavy Higgs bosons with the masses around the EW scale 
(at the quantum loop level) successfully causes a strong first-order PT for the electroweak (EW) symmetry as well as for the scale symmetry. 
Moreover, if we do not impose a $Z_2$ parity on Yukawa sector,
it is possible to have additional Yukawa couplings that induce the charge-conjugation and parity (CP) violation and/or flavor violation, 
which cannot be absorbed into the Cabibbo-Kobayashi-Maskawa matrix. 
As a result, a realistic amount of the baryon asymmetry of Universe (BAU) can be realized~\cite{Fuyuto:2017ewj,Modak:2018csw}  by the EW baryogenesis (EWBG)~\cite{Kuzmin:1985mm} 
(for reviews, see, {\it e.g.}, Refs.~\cite{Quiros:1994dr,Rubakov:1996vz,Funakubo:1996dw,Morrissey:2012db}), 
following the standard sphaleron-freeze out scenario coupled with a chiral/CP violating transport mechanism. 

Along such a cosmological scenario, 
we investigate the chameleon mechanism around EWPT epoch with the realistic setup for parameters in the SI-2HDM supported by the particle phenomenology.
Working on a static analysis of matter sector, which suffices to study the EWPT,
we explicitly compute the trace of energy-momentum tensor arising from the SI-2HDM matter sector
and discuss the effect on the scalaron surrounded by the strong-first order EWPT environment in the early Universe. 
In particular, our main focus will be on the scalaron mass before and after the EWPT. 

We also evaluate the time-evolution of the potential coupled with the SI-2HDM Lagrangian, motivated by the particle physics in the flat background,
which is possible because 
the potential structure does not depend on the background of space-time, 
although several works \cite{Brax:2004qh,Erickcek:2013oma} had studied the time-evolution of the scalar field in the cosmological background. 
We demonstrate that 
the conventional fluid approximation is actually valid after the EWPT,
as far as order-of-magnitude evaluation for the trace of energy-momentum tensor, 
which leads to the scalaron mass and the potential, is concerned.

We thus make a first attempt to evaluate the chameleon mechanism in the early Universe, 
explicitly based on the Lagrangian formalism. 
An indirect generation of non-tensorial gravitational waves induced by 
the strong-first order EWPT is also addressed. 
Though being somewhat specific to the choice of scenarios beyond 
the standard model of particle physics, what we provide in this paper involves 
the essential feature related to the PT for the scale-symmetry breaking, which is 
applicable also to other similar models beyond the standard model. 

This paper is organized as follows.
We briefly introduce the chameleon mechanism in $F(R)$ gravity and explain the dependence on the trace of the energy-momentum tensor in Sec.~\ref{2}. 
We review the SI-2HDM for the EWPT in Sec.~\ref{3}, 
and formulate and compute the trace of the energy-momentum tensor relevant to the EWPT epoch in Sec.~\ref{4}.
In section \ref{5}, we demonstrate the chameleon mechanism in a specific model of $F(R)$ gravity with SI-2HDM,
to find the thermal history of the scalaron field in the EWPT environment. 
Summary and discussions are given in Sec.~\ref{summary}.


\section{$F(R)$ gravity and chameleon mechanism}
\label{2}

In this section, we give a brief review of $F(R)$ gravity and chameleon mechanism for the scalar field.
We also introduce the specific model of $F(R)$ gravity and its properties.

\subsection{Action and Weyl transformation}
\label{2A}

The action of generic $F(R)$ gravity is given as follows:
\begin{align}
S=\frac{1}{2\kappa^{2}} \int d^{4}x \sqrt{-g} F(R) + \int d^{4}x \sqrt{-g} \mathcal{L}_\mathrm{Matter} [g^{\mu \nu}, \Phi]
\label{action1}
\, ,
\end{align}
where $F(R)$ is a function of the Ricci scalar $R$,
and $\kappa^{2} = 8\pi G = 1/M^{2}_\mathrm{pl}$.
$M_\mathrm{pl}$ is the reduced Planck mass 
$\sim 2 \times 10^{18} [\mathrm{GeV}]$.
$\mathcal{L}_\mathrm{Matter}$ denotes the Lagrangian for a matter field $\Phi$,
and the matter field $\Phi$ follows the geodesics of a metric $g_{\mu \nu}$.

The variation with respect to the metric $g_{\mu \nu}$ leads to the equation of motion:
\begin{align}
F_{R}(R) R_{\mu \nu} - \frac{1}{2} F(R) g_{\mu \nu} + (g_{\mu \nu} \Box - \nabla_{\mu} \nabla_{\nu}) F_{R}(R) 
&= \kappa^{2} T_{\mu \nu} (g^{\mu \nu}, \Phi)
\label{jordaneom1}
\, .
\end{align}
Here, $F_{R}(R)$ means the derivative of $F(R)$ with respect to $R$, $F_{R}(R) = \partial_{R} F(R)$, and the energy-momentum tensor $T_{\mu \nu}$ is given by
\begin{align}
T_{\mu \nu}(g^{\mu \nu}, \Phi) 
= \frac{-2}{\sqrt{-g}} \frac{\delta \left(\sqrt{-g} \mathcal{L}_\mathrm{Matter} (g^{\mu \nu}, \Phi) \right)}{\delta g^{\mu \nu}}
\label{EMT}
\, .
\end{align}

We can look into the dynamics of the new scalar field via the Weyl transformation.
It is known that the $F(R)$ gravity is equivalent to the scalar-tensor theory via the Weyl transformation of the metric, which is the frame transformation from the Jordan frame $g_{\mu \nu}$ to the Einstein frame $\tilde{g}_{\mu \nu}$:
\begin{align}
g_{\mu \nu} \rightarrow \tilde{g}_{\mu \nu} 
= 
\mathrm{e}^{2 \sqrt{1/6} \kappa \varphi } g_{\mu \nu} 
\equiv  F_{R} (R) g_{\mu\nu}   
\label{Weyltrans}
\, .
\end{align}
Under the Weyl transformation, the original action Eq.~(\ref{action1}) is transformed as follows:
\begin{align}
S
=&
\frac{1}{2\kappa^{2}} \int d^{4}x \sqrt{-\tilde{g}} \tilde{R}  
\nonumber \\ 
& +
\int d^{4}x \sqrt{-\tilde{g}} \left[ - \frac{1}{2} \tilde{g}^{\mu \nu} (\partial_{\mu} \varphi) (\partial_{\nu} \varphi) - V_{s}(\varphi) \right]
\nonumber \\
& 
+ \int d^{4}x \sqrt{-\tilde{g}} \, \mathrm{e}^{-4 \sqrt{1/6} \kappa \varphi } 
\mathcal{L}_\mathrm{Matter} \left[ \mathrm{e}^{2 \sqrt{1/6} \kappa \varphi }  \tilde{g}^{\mu \nu}, \Phi \right]
\label{action2}
\, .
\end{align}
We call $\varphi(x)$ the scalaron field and define its potential as,
\begin{align}
V_{s}(\varphi) 
\equiv \frac{1}{2\kappa^{2}} \frac{R F_{R}(R) - F(R)}{F^{2}_{R}(R)}
\label{scalaronpotential}
\, .
\end{align}
Note that through the Weyl transformation in Eq.~(\ref{Weyltrans}), 
the Ricci scalar $R$ is given as a function of the scalaron field $\varphi$, such as $R = R(\varphi)$.

By the variation of the action in Eq.~(\ref{action2}) 
with respect to the Einstein frame metric $\tilde{g}_{\mu \nu}$, 
we obtain the Einstein equation with the minimally coupled scalaron field.
The variation with respect to the scalaron field $\varphi$ gives us the equation of motion 
for the scalaron field, 
\begin{align}
\tilde{\Box} \varphi  =&
\frac{\partial V_{s}(\varphi)}{\partial \varphi}
+ \frac{\kappa}{\sqrt{6}} \mathrm{e}^{-4 \sqrt{1/6} \kappa \varphi }  T^{\mu}_{\ \mu}
\label{scalaroneom}
\, .
\end{align}
Note that $T^{\mu}_{\ \mu}$ involves the scalaron field induced by the nonlinear (dilatonic) coupling with matter fields, 
$T^{\mu}_{\ \mu} = T^{\mu}_{\ \mu} (\varphi, \tilde{g}^{\mu \nu}, \Phi )$, from Eqs.~\eqref{EMT} and \eqref{action2}.
From Eq.~(\ref{scalaroneom}),
we define the effective potential of the scalaron field as follows:
\begin{align}
V_{s\, \mathrm{eff.}}(\varphi) = 
V_{s}(\varphi) + \int d\varphi \frac{\kappa}{\sqrt{6}} \mathrm{e}^{-4 \sqrt{1/6} \kappa \varphi }  T^{\mu}_{\ \mu}
\label{scalaron_eff_potential1}
\, .
\end{align}
We note that the effective potential of the scalaron includes the trace of energy-momentum tensor $T^{\mu}_{\ \mu}$. 
In other words, the matter distributions affect the potential structure of the scalaron,
which leads the environment-dependent mass in the scalaron dynamics.
This feature is related to so-called the chameleon mechanism, which we will see later in detail.

In general, $T^{\mu}_{\ \mu}$ has a nontrivial dependence on $\varphi$ 
($T^{\mu}_{\ \mu} = T^{\mu}_{\ \mu} (\varphi)$), 
which comes up from the metric $g^{\mu \nu} = \mathrm{e}^{2 \sqrt{1/6} \kappa \varphi } \tilde{g}^{\mu \nu}$. 
Given a certain type of the $\varphi$-dependence, 
we can find the matter-sector interactions with the scalaron as well as the self-interactions.    
In the previous work by two of authors \cite{Katsuragawa:2017wge}, 
$T^{\mu}_{\ \mu}$ was assumed to be constant in $\varphi$ 
by following the earlier works \cite{Brax:2004qh, Erickcek:2013oma},
to give the simplified formula
\begin{align}
V_{s\, \mathrm{eff.}}(\varphi) = 
V_{s}(\varphi) - \frac{1}{4} \mathrm{e}^{-4 \sqrt{1/6} \kappa \varphi }  T^{\mu}_{\ \mu}
\label{scalaron_eff_potential2}
\, .
\end{align}
Although one can observe the above formula in many literature, 
it is not a precise evaluation but just a modeled one. 
As will turn out later, 
this modeled expression in Eq.(\ref{scalaron_eff_potential2}) is justifiable in the field-theoretical manner,
which actually allows us to apply it directly to the evaluation of chameleon mechanism influenced by a strong-first order EWPT.

\subsection{Chameleon mechanism and energy-momentum tensor}
\label{2B}

Next, we discuss the minimum of the scalaron effective potential and the scalaron mass with the matter effect $T^{\mu}_{\ \mu}$ included.
The first derivative of the scalaron effective potential is written in terms of the $F(R)$ function:
\begin{align}
\frac{\partial V_{s\, \mathrm{eff.}}(\varphi)}{\partial \varphi}
=&
\frac{1}{\sqrt{6} \kappa} \left( \frac{2F(R) - R F_{R}(R) + \kappa^{2}T^{\mu}_{\ \mu} }{F^{2}_{R}(R)} \right)
\label{scalaron_eff_potential_derivative1}
\, .
\end{align}
The minimum of the potential at $\varphi = \varphi_{\min}$ should satisfy the stationary condition
that Eq.~\eqref{scalaron_eff_potential_derivative1} vanishes,
which leads to
\begin{align}
2F(R_{\min}) - R_{\min} F_{R}(R_{\min})  + \kappa^{2} T^{\mu}_{\ \mu} = 0
\label{potentialminimum1}
\, .
\end{align}
Note that $R_{\min}$ is related to $\varphi_{\min}$ through the Weyl transformation $\mathrm{e}^{2 \sqrt{1/6} \kappa \varphi_{\min} } =  F_{R} (R_{\min}) $.
Moreover, the square of scalaron mass $m_{\varphi}$ is defined as the value of the second derivative of the effective potential at the minimum.
The second derivative of the effective potential is evaluated as follows: 
\begin{align}
\frac{\partial^{2} V_{s\, \mathrm{eff.}}(\varphi)}{\partial \varphi^{2}}
&=
\frac{1}{3F_{RR}(R)}  \left[ 1 + \frac{ R F_{RR}(R)}{F_{R}(R)}  
- \frac{ 2 \left( 2 F(R) + \kappa^{2}T^{\mu}_{\ \mu} \right) F_{RR}(R)}{F^{2}_{R}(R)} 
\right]
\label{scalaron_eff_potential_derivative2}
\, .
\end{align}
Substituting Eq.~(\ref{potentialminimum1}) into Eq,~(\ref{scalaron_eff_potential_derivative2}), we obtain
\begin{align}
m^{2}_{\varphi} (T^{\mu}_{\ \mu})
&=
\frac{1}{3F_{RR}(R_{\min})}  \left(1 - \frac{ R_{\min} F_{RR}(R_{\min})}{F_{R}(R_{\min})} \right) 
\, .\label{mass:formula1}
\end{align}
Note that since the stationary condition Eq.~(\ref{potentialminimum1}) determines $\varphi_{\min}$ or $R_{\min}$,  
the scalaron mass changes according to the trace of the energy-momentum tensor $T^{\mu}_{\ \mu}$.

As Eqs.~\eqref{scalaron_eff_potential1} and \eqref{mass:formula1} show,
the effective potential and mass of the scalaron depend on the trace of the energy-momentum tensor $T^{\mu}_{\ \mu}$,
which is a key part of the chameleon mechanism.
Now, we consider more on the construction of the energy-momentum tensor.
In the context of the cosmology and astrophysics, 
various literature has employed the fluid description to express the environment surrounding the scalaron field (for example, see \cite{Capozziello:2018wul}),
and the trace of the energy-momentum tensor is computed as
\begin{align}
T^{\mu}_{\ \mu} = - (\rho - 3p)
\end{align}
where $\rho$ and $p$ are the energy density and pressure of the fluid.

The simplest case is the pressure-less dust $T^{\mu}_{\ \mu} = -\rho$,
which is a good approximation to describe the matters in the current Universe, 
and the chameleon mechanism is controlled by the energy density $\rho$.
For instance, Eq.~\eqref{scalaroneom} is reduced to 
\begin{align}
\tilde{\Box} \varphi 
= \frac{\partial V_{s}(\varphi)}{\partial \varphi}
- \frac{\kappa}{\sqrt{6}} \rho \mathrm{e}^{-4 \sqrt{1/6} \kappa \varphi }  
\label{scalaron_eff_potential2-1}
\, .
\end{align}
If one can design the $F(R)$ function so that the scalaron mass becomes large enough in the high-density region,
the scalaron is screened because of its heavy mass.
This feature is called chameleon mechanism which is one of the screening mechanism in the modified gravity.
As a consequence, for example, the scalaron becomes heavy in the Solar System,
where the scalaron field is screened, and the $F(R)$ gravity can be relevant to the observations.
On the other hand, in the low-energy density environment, that is, on the cosmological scale,
the scalaron field becomes dynamical dark energy.

Here, we emphasize that the trace of the energy-momentum tensor $T^{\mu}_{\ \mu}$ is not necessarily evaluated in the framework of the conventional fluid approximation,
and thus, one can utilize the other frameworks for the different purpose.
For instance, it is natural that
one should compute $T^{\mu}_{\ \mu}$ in framework of the quantum field theory
in the early and high-energetic epoch of the Universe.
Because we are interested in the EWPT in the early Universe,
we introduce the suitable model beyond the standard model of particle physics as the ingredients for the energy-momentum tensor.
In the rest of paper, we formulate the trace of the energy-momentum tensor with the field-theoretical techniques
and compute $T^{\mu}_{\ \mu}$ for the model of our interest. 
Then, we substitute it into Eqs.~\eqref{scalaron_eff_potential1} and \eqref{mass:formula1}
to study the scalaron physics and related chameleon mechanism in the EWPT epoch.

\section{EWPT in a SI-2HDM}
\label{3}

In this section, 
we consider a scale-invariant embedding for a scenario beyond the standard model of particle physics 
and compute the effective potential of the matter sector
to discuss how the scalaron is affected by the EWPT environment in the early Universe.
As a reference scenario beyond the standard model,
we shall take a class of general 2HDM which yields a realistic cosmic history with realization of the right amount of BAU accompanied with a desired strong-first order EWPT~\cite{Fuyuto:2015vna, Fuyuto:2017ewj, Modak:2018csw}. 
In the present study, we shall apply such an EWBG scenario by extending it to be scale-invariant (general SI-2HDM).

\subsection{A class of general SI-2HDM}
\label{3A}

We begin with introducing a general SI-2HDM, defined by the following Lagrangian:
\begin{align}
\mathcal{L} = \mathcal{L}_{\mathrm{2HDM|w/o} \ V_{0}} - V_{0} \left( \Phi_{1}, \Phi_{2} \right)
\, ,
\end{align}
where $V_0$ is the tree-level Higgs potential and the two Higgs doublets $(\Phi_{1}, \Phi_{2})$ are parametrized in terms of the fluctuation fields in the broken phase as
\begin{align}
\Phi_{i} = 
\left(
\begin{array}{c}
      \phi_{i}^{+}  \\
      \frac{1}{\sqrt{2}} \left[ v_{i} + h_{i}(x) + ia_{i }\right]
    \end{array}
  \right)
\, , 
\quad \mbox{for} \quad i=1,2
\, .\label{general:HD}
\end{align}
Their vacuum expectation values (VEVs) are characterized as $v_{1}=v \cos \beta$ and $v_{2}=v \sin \beta$, 
in which $v \simeq 246 [\mathrm{GeV}]$.

Hereafter, we work in the Higgs-Georgi basis, where all the Nambu-Goldstone (NG) bosons do not show up in the physical Higgs spectra,
and the only one Higgs doublet $\Phi_{1}$ acquires the VEV $v \simeq 246 [\mathrm{GeV}]$. 
The two Higgs doublet fields $(\Phi_1,\Phi_2)$ in Eq.(\ref{general:HD}) are transformed 
by the orthogonal-basis rotation with the angle $\beta$ 
into $(H_1,H_2)$ as follows:
\begin{align}
H_1 & = 
\left(
\begin{array}{c}
      G^+ \\
      \frac{1}{\sqrt{2}} \left[ v + h_1^\prime + i G^0 \right]
    \end{array}
  \right)
\nonumber \,, \\ 
H_2 &= 
\left(
\begin{array}{c}
      H^+ \\
      \frac{1}{\sqrt{2}} \left[ v + h_2^\prime + i A \right]
    \end{array}
  \right)
\, , \nonumber \\ 
\left( 
\begin{array}{c} 
h_1^\prime \\ 
h_2^\prime 
\end{array}
\right) 
&  =
\left( 
\begin{array}{cc} 
c_\beta  & s_\beta \\ 
- s_\beta & c_\beta  
\end{array}
\right) 
\left( 
\begin{array}{c} 
h_1  \\ 
h_2  
\end{array}
\right) 
\, , 
\end{align}
where $c_\beta \equiv \cos \beta$ and $s_\beta \equiv \sin \beta$.
$G^{\pm,0}$ denote the NG boson fields to be eaten by 
$W^\pm$ and $Z$. 
The neutral Higgs fields $(h_1, h_2)$ will be  
further orthogonally rotated by an angle $\alpha$ to be mass eigenstate fields $(h, H)$ 
in a similar manner. 

The tree-level Higgs potential in the Higgs-Georgi basis is defined as
\begin{align} 
V_{0} \left(H_{1}, H_{2} \right) 
= &
\frac{\lambda_{1}}{2} \left( H_1^{\dagger} H_1 \right)^{2}
+ \frac{\lambda_{2}}{2} \left( H_2^{\dagger} H_2 \right)^{2}
+ \lambda_{3} \left( H_1^{\dagger} H_1 \right) \left( H_2^{\dagger} H_2 \right)
+ \lambda_{4} \left( H_1^{\dagger} H_2 \right) \left( H_2^{\dagger} H_1 \right)
\nonumber \\ 
& 
+ \left \{ \frac{\lambda_{5}}{2} \left( H_1^{\dagger} H_2 \right)^{2}
+ \left[ \lambda_{6} \left( H_1^{\dagger} H_1 \right)
+ \lambda_{7} \left( H_2^{\dagger} H_2 \right) \right] \left( H_1^{\dagger} H_2 \right)
+ \mbox{h.c.} \right \}
\label{SI-2HDM_potential_1}
\, ,
\end{align}
from which the tadpole conditions at tree-level are found to be
\begin{align}
\frac{\lambda_{1}}{2} v^3 = 0 \, , \ 
\quad 
\frac{\lambda_{6}}{2} v^3 = 0 \, .
\end{align}
Therefore, for nonzero $v$, we find
\begin{align}
\lambda_{1} = \lambda_{6} = 0, \, , \
\quad 
V_{0}(v)=0 \, .
\end{align} 

The masses of the charged Higgs $H^{\pm}$, CP-odd Higgs $A$, and CP-even Higgs $\left( h, H \right)$
are evaluated as
\begin{align}
m_{H^{\pm}}^{2} &= \frac{\lambda_{3}}{2} v^2
\, , \qquad 
m_{A}^{2} = \frac{1}{2} \left( \lambda_{3} + \lambda_{4} - \lambda_{5} \right) v^2 \, , 
\nonumber \\
m_{\mathrm{even}}^{2} &\equiv
\left(
\begin{array}{cc}
      m_{h}^{2} & 0  \\
      0 & m_{H}^{2}
    \end{array}
  \right)
=
\left(
\begin{array}{cc}
      0 & 0  \\
      0 & \frac{1}{2} \left( \lambda_{3} + \lambda_{4} + \lambda_{5} \right) v^2
    \end{array}
  \right)
\, .
\end{align}
Here, the massless neutral Higgs $(h)$ in the mass matrix $m_{\rm even}^2$ is called ``{\it scalon}''~\cite{Gildener:1976ih} (not confused with scalaron), which arises as the consequence of the classical-scale invariance in the present model.
This massless {\it scalon} ensures the existence of a flat direction in the potential. 
We will investigate the EW symmetry breaking along this flat direction 
by assuming the {\it scalon}-Higgs mass to be zero at some renormalization scale.

Regarding the choice of the potential parameters, 
we further assume the custodial symmetric limit to protect the possibly sizable contribution to the $\rho$ parameter, 
which is set as
\begin{align}
m_{H^{\pm}}^{2} = m_{A}^{2} 
\, ,
\end{align}
and then, we obtain
\begin{align}
\lambda_{4} = \lambda_{5} = \frac{m_{H}^{2} - m_{A}^{2}}{v^{2}}
\, .
\end{align}
In addition, we take $m_{H} = m_{A}$ for the benchmark point in addressing the EWPT and baryogenesis. 
In this case, we have 
\begin{align}
\lambda_{3} = \frac{2m_{H}^{2}}{v^{2}} \, , \ 
\lambda_{4} = \lambda_{5} = 0 \, .
\label{lambda3-4-5}
\end{align}
Finally, for the benchmark point where $m_{H} = m_{A} = m_{H^{\pm}}$ and $t_{\beta} \equiv \tan \beta = 1$,
the Higgs potential in the present model 
is controlled only by the parameters $\lambda_{3}$ and $\lambda_{7}$:
\begin{align}
V_{0}(\phi) 
&= \frac{\lambda_{3}}{2} \left(2\phi h + \phi^{2}\right) \left( H^{+} H^{-} + \frac{1}{2} A^{2} + \frac{1}{2} H^{2} \right)
- \lambda_{7}\phi H \left( H^{+} H^{-} + \frac{1}{2} A^{2} + \frac{1}{2} H^{2} \right)
\label{SI-2HDM_potential_2}
\, ,
\end{align}
where $\phi = \sqrt{\phi_{1}^{2} + \phi_{2}^{2}}$ with $\phi_{i}$ being the constant background fields of the two Higgs doublets. 

As long as the effective potential for the background field $\phi$ is evaluated at the one-loop level, 
the second term with the coupling $\lambda_7$ in Eq.~\eqref{SI-2HDM_potential_2} does not contribute 
to the effective potential,
but only the first term with the coupling $\lambda_3$ does.
It will be replaced by the heavy Higgs mass coupling Eq.~\eqref{lambda3-4-5}
and reduced to be the field-dependent (common) masses for the $H, H^\pm$ and $A$ 
as will be seen in Eq.(\ref{FDM}).
Thus, we can straightforwardly quote the result in the EWPT 
as well as the sphaleron freeze-out condition in \cite{Fuyuto:2015vna}
where the analyzed model has been set up in the scale-invariant limit, 
but not generic due to the requirement of a $Z_2$ symmetry among the two Higgs doublet fields. 
In particular, the numerical values listed in Table 1 of the reference can be directly applied even in the general 2HDM-setup, 
as we will see them in the next subsection. 

Note that the only one exception is on estimation for the cutoff scale $\Lambda$ 
regarding the present general SI-2HDM: 
one can compute the one-loop renormalization group equations for the potential couplings
(for the explicit expressions, see \cite{Branco:2011iw}). 
The straightforward one-loop computation tells us 
that the present model has a Landau pole ($\Lambda_{\rm LP}$) at the scale $\simeq 8.8 [\mathrm{TeV}]$
which is regarded as a cutoff scale up to which the present model is valid. 
This computation is based on the Higgs-Georgi basis with the massless Higgs and other related inputs
which are used in the later subsection.
$\tilde{\mu}$ is determined by other inputs (See, Eq. (33)).
$\Lambda_{\rm LP} \simeq 8.8 [\mathrm{TeV}]$ in the present model
is somewhat larger than that in the SI-2HDM with the $Z_2$ symmetry 
imposed ($\Lambda_{\rm LP} \simeq 6.3 [\mathrm{TeV}]$)~\cite{Fuyuto:2015vna}.

\subsection{The one-loop effective potential at zero temperature: EW symmetry breaking}
\label{3B}

We follow the Gildener-Weinberg method~\cite{Gildener:1976ih} to compute the one-loop effective potential at zero temperature, 
$V_{1}(\phi)$,  along the flat direction at tree-level. 
In the  Gildener-Weinberg method, 
the NG bosons are exactly massless along the flat direction, 
hence they do not contribute to the one-loop effective Higgs potential, 
making the potential gauge invariant~\footnote{
At finite temperature, however, the NG bosons get thermal masses, rendering the effective potential gauge dependent after thermal resummation~\cite{Chiang:2017zbz} (see also Ref.~\cite{Patel:2011th}). Since there is no satisfactory
gauge-invariant perturbative calculation method at present, we do not pursue this issue and take Landau gauge in our numerical analysis. In this gauge, the NG contributions are $\phi$ independent at leading order in resummed perturbation theory so that they do not appear in the following calculations. 
}.
Thus we can readily compute the one-loop effective potential 
by using the tree-level relation 
$\lambda_{3} = 2m_{H}^2/v^2$ in Eq.(\ref{lambda3-4-5})
and the usual Yukawa and gauge interaction terms in the standard model,
\begin{align}
V_{1} (\phi) 
= \sum_{i=H,A,H^{\pm},W^{\pm},Z,t,b} n_{i} \frac{\tilde{m}(\phi)_{i}^{4}}{64 \pi^2} 
\left( \log \frac{\tilde{m}_{i}^{2}(\phi)}{\tilde{\mu}^2}  - c_{i} \right)
\label{SI-2HDM_one-loop_potential}
\, ,
\end{align}
where 
$n_{i}$ stands for the degree of freedom for each particle 
(n.b., a minus sign appears for fermion loops):
$n_{H} = n_{A} = 1$, $n_{H^{\pm}} = 2$, $n_{W^{\pm}} = 6$, $n_{Z} = 3$, $n_{t} = n_{b} = -12$, 
and $c_{i} = 3/2 \ (5/6)$ for scalars and fermions (gauge bosons). 
They come in the potential due to the ($\overline{\text{MS}}$) renormalization procedure at one-loop level. 
In Eq.(\ref{SI-2HDM_one-loop_potential}), 
the $\tilde{m}(\phi)$ denotes the field-dependent masses for the heavy Higgses $(H, A, H^{\pm})$ and the standard-model particles
(with being selected to be relatively heavy ones with the larger couplings to the $\phi$, such as $W^{\pm}, Z, t, b$), 
which are defined by 
\begin{align} 
\tilde{m}_{i}^{2} = m_{i}^{2} \frac{\phi^{2}}{ v^{2}}  
\,. \label{FDM}
\end{align}

Since the VEV $v$ does not develop at the tree-level as the consequence of the classically scale-invariant setup, 
$v$ emerges through the renormalization scale $(\tilde{\mu})$ at the one-loop reflecting the dimensional transmutation,
which is usually called radiative-EW breaking-mechanism. 
The tadpole condition at the one-loop level $\partial V_{1} (\phi) / \partial \phi = 0 |_{\phi=v}$ leads to 
\begin{align}
v^{2} &= \tilde{\mu}^2 \exp \left[- \frac{1}{2} - \frac{A}{B} \right] 
\, , \nonumber \\
A &= \sum_{i=H,A,H^{\pm},W^{\pm},Z,t,b} n_{i} \frac{\tilde{m}_{i}^{4}(\phi)}{64 \pi^2 v^{4}} 
\left( \log \frac{\tilde{m}_{i}^{2}(\phi)}{v^2}  - c_{i} \right)
\, , \nonumber \\
B &=  \sum_{i=H,A,H^{\pm},W^{\pm},Z,t,b} n_{i} \frac{\tilde{m}_{i}^{4}(\phi)}{64 \pi^2 v^{4}}
\, . \label{vmu}
\end{align}
Then the vacuum energy becomes 
\begin{align} 
V_1(v) = - \frac{B}{2} v^4 
\, ,
\end{align}
which has to be negative (i.e. $B>0$ since $V_0(v)=0$ along the flat direction)  
so as to realize the EW breaking at the true vacuum. 
Eliminating the renormalization scale $\tilde{\mu}$ by using Eq.~\eqref{vmu}, 
the one-loop effective potential takes the form 
\begin{align} 
V_1(\phi) = B \phi^4 \left( \log \frac{\phi^2}{v^2}  - \frac{1}{2} \right)
\,, 
\end{align}
and the 125 GeV Higgs mass is thus 
radiatively generated as follows:
\begin{align}
(125 [\mathrm{GeV}])^2 = m_{h}^{2} 
= \left. \frac{\partial^2 V_{1} (\phi)}{\partial \phi^2} 
\right |_{\phi=v} = 8Bv^2 
\, . 
\end{align}

\subsection{The one-loop effective potential at finite temperature: EWPT}
\label{3C}

Including the finite temperature effect via the imaginary-time formalism 
and applying the resummation prescription \cite{Parwani:1991gq,Buchmuller:1992rs,Chiku:1998kd,Funakubo:2012qc},
the one-loop potential Eq.~\eqref{SI-2HDM_one-loop_potential} receives the corrections,
and we obtain the effective potential $V_{h\, \mathrm{eff.}} (\phi, T)$:
\begin{align}
&V_{h\, \mathrm{eff.}} (\phi, T) \nonumber \\ 
=  & 
\sum_{\substack{i=H,A,H^{\pm}, W_{L,T}^{\pm},\\ Z_{L,T}, \gamma_L,t,b}} n_{i} 
\left [
\frac{\tilde{M}_{i}^{4}(\phi,T)}{64 \pi^2} 
\left( \log \frac{\tilde{M}_{i}^{2}(\phi,T)}{\tilde{\mu}^2}  - c_{i} \right)  
+ 
\frac{T^{4}}{2\pi^{2}} I_{B,F} \left( \frac{\tilde{M}_{i}^{2}(\phi,T)}{T^2} \right)
\right ]
\label{SI-2HDM_effective_potential}
\, , 
\end{align}
where $n_{W_{L(T)}}=2(4)$, $n_{Z_{L(T)}}=1(2)$, $c_{V_{L(T)}}=3/2(1/2)~(V=W, Z)$.
The field-dependent masses at the finite temperature $\tilde{M}_{i}^{2} (\phi, T)$ are given by
\begin{align}
& \tilde{M}_{H,A,H^\pm}^{2}(\phi,T) 
= \tilde{m}_{H,A,H^\pm}^{2} (\phi) + \Pi_{H,A,H^\pm} (T)\,, \\ 
& \tilde{M}_{W_L}^2(\phi, T) = \tilde{m}_W^2(\phi)+\Pi_W(T)\,,\\
&\tilde{M}_{Z_L, \gamma_L}^2(\phi, T)
=\frac{1}{2}
\left [
\frac{1}{4}(g_2^2+g_1^2)\phi^2 + \Pi_W(T) + \Pi_B(T) 
\right.
\nonumber\\
& \left. \qquad \qquad \qquad \qquad \qquad
	\pm \sqrt{\left(\frac{1}{4}(g_2^2-g_1^2)\phi^2+\Pi_W(T)-\Pi_B(T)\right)^2
	+\frac{g_2^2g_1^2}{4}\phi^4}
\right ]
\, ,
\end{align}
and for each field~\cite{Carrington:1991hz},
\begin{align}
\Pi_{H, A, H^{\pm}} (T) &= \frac{T^{2}}{12 v^{2}} 
\left ( 6m_{W}^{2} + 3m_{Z}^{2} + 4m_{H}^{2} 
\right. \nonumber \\ 
& \left. \qquad +  6 m_{t}^{2} + 6m_{b}^{2} \right ) 
\, , \\
\Pi_W(T) &= 2g_2^2T^2\,, \\
\Pi_B(T) &= 2g_1^2T^2\,,
\end{align}
where $g_2$ and $g_1$ are the $\text{SU(2)}_L$ and $\text{U(1)}_Y$ couplings, respectively.
For the other species, $\tilde{M}_i^2(\phi, T) =\tilde{m}_i^2(\phi)$.
And, $ I_{B,F} ( \tilde{M}_{i}^{2}(\phi,T)/T^2 )$ is defined by
\begin{align}
I_{B,F} (a^{2})&=  \int^{\infty}_{0} dx x^{2} \log \left( 1 \mp \mathrm{e}^{-\sqrt{x^{2} + a^{2}}} \right)
\, ,
\end{align}
where the minus sign is applied for bosons and the plus one for fermions.
For the numerical evaluations of $I_{B,F}(a^2)$ and their derivatives with respect to $a^2$, we employ fitting functions used in Ref.~\cite{Funakubo:2009eg}. Their errors are small enough for our purpose.

With the effective potential in Eq.~\eqref{SI-2HDM_effective_potential},
we can analyze the EWPT and sphaleron freeze-out 
after we normalize the effective potential to be 0 at $\phi=0$, 
(i.e., making a shift $V(\phi, T) \to V(\phi, T) - V(\phi=0, T)$). 
For successful EWBG, one has to satisfy $v/T>\xi_{\text{sph}}(T)$ at a transition temperature $T$ (described below),
where $\xi_{\text{sph}}(T)$ predominantly depends on sphaleron energy~\cite{Manton:1983nd,Klinkhamer:1984di}.
We take $m_{H}^{2} = m_{A}^{2} = m_{H^{\pm}}^{2} (= 382 [\mathrm{GeV}])^2$ \cite{Fuyuto:2015vna}) as the benchmark point. 
It turns out that all the results are the same as given in 
Table 1 of \cite{Fuyuto:2015vna} 
except for the cutoff scale around $10$ [TeV], as noted in the previous subsection.

Therefore, we can directly quote the successful 
benchmark parameters relevant to the strong first-order PT 
at the critical temperature $(T_C)$ and the nucleation temperature for the 
EW-broken phase bubble $(T_N)$~\cite{Fuyuto:2015vna}:  
\begin{align}
v_{C}/T_{C} &= 211 [\mathrm{GeV}]/91.5 [\mathrm{GeV}] = 2.31
\,, \nonumber \\  
\xi_{\mathrm{sph}}(T_{C})  
&= 1.23
\, , 
\nonumber \\
v_{N}/T_{N} &= 229 [\mathrm{GeV}]/77.8 [\mathrm{GeV}] = 2.94
\,, \nonumber \\  
\xi_{\mathrm{sph}}(T_{N}) 
&= 1.20 
\,, \nonumber \\
E_{\mathrm{cb}}(T_{N})/T_N 
&= 151.7 \, ,
\label{EWPT:parameters}
\end{align}
for $m_{h} = 125 [\mathrm{GeV}]$, $m_{H}=m_{A}=m_{H^{\pm}} = 382[\mathrm{GeV}]$, and $t_{\beta} = 1$.
$v_{C/N}$ is the Higgs VEV at $T_{C(N)}$,  
$\xi_{\mathrm{sph}}(T_{C(N)})$ denotes the related sphaleron decoupling parameter,
and $E_{\rm cb}(T_N)$ represents the energy of the critical bubble (three-dimensional bounce action)~\cite{Linde:1981zj}. 
In calculating $\xi_{\text{sph}}$, thermal effects on the sphaleron configuration are also taken into account. This is the reason why $\xi_{\text{sph}}$ is slightly greater than a conventional rough criterion of $\xi_{\text{sph}}=1$.

The parameters listed in Eq.~\eqref{EWPT:parameters} makes it possible to 
accumulate the realistic amount of BAU
by introducing a moderate size of extra CP-violating Yukawa couplings in the top-charm sector~\cite{Fuyuto:2017ewj} 
or bottom-strange sector~\cite{Modak:2018csw}.
Actually, 
the benchmark value for the heavy Higgs mass ($382 [\mathrm{GeV}]$) 
is somewhat smaller than those adopted in~\cite{Fuyuto:2017ewj} ($500 [\mathrm{GeV}]$) 
and~\cite{Modak:2018csw} ($600 [\mathrm{GeV}]$), 
as well as the related quantities such as $T_{C}, T_{N}$ and so on. 
In evaluating the chiral/CP-violating transport process, 
however, 
such a small range of the mass difference
will not give a significant effect on the (coupled) diffusion rates and the thermal decay rates 
unless extra colored particles are present.

Therefore, the generated BAU in the SI-2HDM is expected to be the same order in magnitude as that estimated in \cite{Fuyuto:2017ewj, Modak:2018csw} (for theoretical uncertainties of the BAU calculation, see \cite{Fuyuto:2017ewj, Modak:2018csw}). 
However, one crucial difference is that
our scenarios do not suffer from any severe experimental constraints,
such as the electric dipole moment of electron 
whose upper limit has been improved down to $1.1\times 10^{-29}~[e~\text{cm}]$ by ACME Collaboration~\cite{Andreev:2018ayy},
since so-called {\it alignment limit}, $\sin(\beta-\alpha)\to1$, is naturally realized in the current model.

\section{ Static $T^{\mu}_{\ \mu}$ in the EWPT}
\label{4}

In the previous section, 
we have introduced the SI-2HDM to describe the EWPT in the matter sector.
Now, we consider the method to evaluate the trace of energy-momentum tensor in a way relevant to such a PT environment.
As will be addressed below, the key point is to note that  
since the evolution of the vacuum state (as well as the mass spectrum) during the EWPT can be described by the free energy, namely an effective potential, 
the variation of the trace of energy-momentum tensor around that epoch can be static.  
Such a static trace of the energy-momentum tensor, $T^{\mu}_{\ \mu} |_{\rm static}$,  can be computed in the field theory,
and we will apply this procedure to the SI-2HDM.
We also discuss the comparison of the (static) $T^{\mu}_{\ \mu}$s derived from the proposed field-theoretical approach and the conventional fluid approximation.

\subsection{The static $T^{\mu}_{\ \mu}$ and decoupling of scalaron} 
\label{4A}

To achieve the complete evaluation of the trace of the energy-momentum tensor, 
we have to take into account the nonlinear coupling between the scalaron and matter fields,
which is technically hard to accomplish.
Noting characteristic features which we are generically faced with in evaluation of the trace of energy-momentum tensor,
we demonstrate how the complexity of the issue can be relaxed in the present case 
we mainly concern about.

First of all, we note that in contrast to the fluid picture where the couplings to scalaron are implicit,  
we have to pay our attention to the interactions between the scalaron and the target-matter sector induced by the Weyl transformation: 
The Lagrangian for the matter sector can be read off 
from Eq.~\eqref{action2} as follows:
\begin{align}
S_{\mathrm{Matter}}
=&
\int d^{4}x \sqrt{-\tilde{g}} \, \mathrm{e}^{-4 \sqrt{1/6} \kappa \varphi } 
\mathcal{L}_\mathrm{Matter} \left[\mathrm{e}^{2 \sqrt{1/6} \kappa \varphi }  \tilde{g}^{\mu \nu}, \Phi \right]
\, .
\end{align}
That is, in addition to the overall $\exp[-4 \sqrt{1/6} \kappa \varphi]$,
we need to consider couplings between the scalaron and matter sector induced by the metric through the kinetic terms, 
which in general causes the non-trivial $\varphi$-dependence of the $T^{\mu}_{\ \mu}$ as in Eq.~\eqref{scalaron_eff_potential1}.

However, it is actually not such a complicated case as far as the PT epoch is concerned:  
Because one can consider that the evolution of the vacuum (including the mass spectra) during the EWPT is (quasi-) static process,
the trace of the energy-momentum tensor can be identified with the static one,
\begin{align}
T^{\mu}_{\ \mu} = \left. T^{\mu}_{\ \mu} \right|_{\rm static} 
\qquad \textrm{in the EWPT epoch}  \, .
\end{align}
By using this $T^{\mu}_{\ \mu} |_{\rm static}$,
we can safely ignore the nontrivial $\varphi$-dependence in the trace of the energy-momentum tensor $T^{\mu}_{\ \mu}$ arising from the kinetic terms. 
The static trace of energy-momentum tensor can be  
simply evaluated
by the scale transformation of the effective potential for the matter sector: 
\begin{align}
\left. T^{\mu}_{\ \mu} \right|_{\rm static} 
= - \delta_{D} V_{\mathrm{Matter}} 
\label{dilation_lagrangian}
\end{align}
up to total derivative. 
$\delta_{D}$ represents the operation of the infinitesimal scale (or dilatation) transformation. 
Note that Eq.~\eqref{dilation_lagrangian} implies
that the static
trace of the energy-momentum tensor does not depend on the metric. 
Moreover, in this case, 
the original expression for the effective potential of the scalaron in Eq.~\eqref{scalaron_eff_potential1} is reduced to 
the modeled one in Eq.~\eqref{scalaron_eff_potential2}.

Although the intrinsic $\varphi$-dependence arising from the Wely transformation is ignored,
we still have the overall $\exp[-4 \sqrt{1/6} \kappa \varphi]$,
which leads to the interactions between the scalaron and potential terms in the matter sector.
As the second step, we introduce a perturbative picture in the scalaron field discussed in \cite{Katsuragawa:2016yir}.
When we consider the fluctuation mode of scalaron field
\begin{align}
\kappa \varphi \rightarrow \kappa \varphi_{\min} + \kappa \varphi
\, , \ 
|\kappa \varphi| \ll 1
\, , 
\end{align}
nonlinear exponential form of scalaron can be reduced into the polynomial form:
\begin{align}
\mathrm{e}^{Q\kappa \varphi} \rightarrow& \mathrm{e}^{Q\kappa \varphi_{\min}} \mathrm{e}^{Q\kappa \varphi}
\nonumber \\
=& \mathrm{e}^{Q\kappa \varphi_{\min}} \left( 1 + Q\kappa \varphi + \cdots \right)
\, ,
\end{align}
where $Q$ is an arbitrary coefficient.
$\mathrm{exp}[Q \kappa \varphi_{\min}]$ corresponds to the difference between the Jordan and Einstein frames.

One can find that the leading order does not include interaction with the scalaron
and that the contributions from the scalaron couplings appear at the loop-order.
Because such interaction terms are suppressed by the Planck mass scale $\kappa \propto 1/M_{\mathrm{pl}}$,
we can neglect them at the EWPT scale in which we work.
Thus, the trace of the energy-momentum tensor can be evaluated
only with the original matter sector and without including the scalaron, up to the frame difference.
As we will see in the final section, we will find $\mathrm{exp}[Q \kappa \varphi_{\min}] \sim 1$,
and we can also ignore the frame difference.
We note that 
thermal loop corrections by the SI-2HDM are included in itself in term of the effective potential $V_{h\, \mathrm{eff.}} (\phi, T) $,
which does not include the loop corrections by the scalaron.

Thereby, Eq.~\eqref{scalaron_eff_potential2} is actually applicable directly to the case 
in which the scalaron acts as a background field overall coupled to matter-sector dynamics, 
as has been discussed in the present analysis~\footnote{
It should be noted that 
Eq.~\eqref{dilation_lagrangian} works only in  
the static analysis on the $T^{\mu}_{\ \mu}$ to be reliable in an environment such as 
PTs, as in our concern in the present study. 
When the matter sector is evaluated as a dynamical bulk, 
in general, loop corrections yield a nontrivial dependence on the metric coming 
along with derivatives, to show up in the dynamical effective action to be 
transferred into the trace of energy-momentum tensor, 
which accompanies the non-minimal couplings with scalaron fields.}.
Thus, we can make the scalaron completely decoupled from the dynamics of the target matter sector, 
which allows us to compute the trace of the energy-momentum tensor as we do in the Jordan frame.
We also note that
the correspondence between the fluid prescription and field theory is not so straightforward
because there are many theoretical difficulties to reproduce the fluid picture starting from the field-theoretical viewpoint
although we have derived Eq.~\eqref{scalaron_eff_potential2} under the several relevant assumptions.

\subsection{ The static $T^{\mu}_{\ \mu}$ around the EWPT epoch in the SI-2HDM}
\label{4B}

By incorporating the general SI-2HDM in the previous section into the targeted matter Lagrangian,
we can evaluate the $T^{\mu}_{\ \mu}$ as
\begin{align}
T^{\mu}_{\ \mu}
=
\left. T^{\mu}_{\ \mu} \right|_{\rm static}
= - \delta^{\phi}_{D} \left[ V_{h\, \mathrm{eff}} (\phi, T) \right] |_{\phi = v(T)}
\label{thermal_average_energy-momentum_tensor}
\, ,
\end{align}
where $V_{h\, \mathrm{eff}} (\phi, T)$ is given in Eq.~\eqref{SI-2HDM_effective_potential},
and $\delta^{\phi}_{D}$ expresses the infinitesimal operator $\delta_{D}$ with respect to the Higgs field $\phi$, 
which is constant in the space-time: 
$\delta^{\phi}_{D} \varphi = \varphi$~\footnote{
Here, we note again that
the effective potential $V_{h\, \mathrm{eff}} (\phi, T)$ does not involve the scalaron field
due to the metric-independence.
When we handle the effective action,
the non-minimal couplings between the scalaron and fields in SI-2HDM are generated.
However, Eq.~\eqref{thermal_average_energy-momentum_tensor} is valid 
as far as the effect from the EWPT environment on the scalaron is concerned.}. 
Eq.~\eqref{thermal_average_energy-momentum_tensor} can be computed as follows:  
\begin{align}
&\delta^{\phi}_{D} \left[ V_{h\, \mathrm{eff}} (\phi, T) \right] |_{\phi = v(T)}
\nonumber \\
=&
\sum_{i} n_{i} \tilde{m}_{i}^{2}(\phi)
\left. \left[ \frac{\tilde{M}_{i}^2(\phi,T)}{16 \pi^2} 
\left( \log \frac{\tilde{M}^2_i (\phi,T)}{\tilde{\mu}^2}  - c_{i} + \frac{1}{2} \right) 
+ 
\frac{T^2}{\pi^{2}} 
I_{B,F}^\prime \left( \frac{\tilde{M}_{i}^{2}(\phi,T)}{T^2} \right)
 \right] \right |_{\phi = v(T)}
\label{thermal_average_energy-momentum_tensor2}
\, .
\end{align} 
where $I_{B,F}^\prime(a^2) \equiv \frac{\partial}{\partial a^2} 
I_{B,F}(a^2)$.
Here, the temperature dependence of the Higgs VEV $v(T)$ is completely determined by the potential analysis in the
previous section.

The scalaron $\varphi$ couples to the $T^{\mu}_{\ \mu}$
as in Eq.~\eqref{scalaron_eff_potential2} (with the $T^{\mu}_{\ \mu}$ replaced by 
the static one in Eq.(\ref{thermal_average_energy-momentum_tensor})), 
with the dilatation variance given in Eq.~\eqref{thermal_average_energy-momentum_tensor2}, 
so that the scalaron dynamics gets significantly affected by the 
matter sector, the SI-2HDM, after the EWPT at $T=T_C$. 
There, 
the explicit scale symmetry breaking is provided by introduction of the renormalization scale $\tilde{\mu}$ 
as well as the temperature $T$ transported from the one-loop effective potential of the Higgs $\phi$ field, 
as clearly seen from Eq.(\ref{thermal_average_energy-momentum_tensor2}). 
To be more explicit, 
we expand the Higgs field $\phi(T)$ in Eq.(\ref{thermal_average_energy-momentum_tensor2}) 
around the VEV $v(T)$ ($T\le T_C$) 
by introducing the fluctuating field $\phi$ via a shift $v \to v + \phi$.
We can read off the coupling terms between the Higgs $\phi$ and the scalaron $\varphi$ 
in the scalaron effective potential 
(in Eq.~\eqref{scalaron_eff_potential2} 
with the $T^{\mu}_{\ \mu}$ replaced by the static one in 
Eq.(\ref{thermal_average_energy-momentum_tensor}), 
together with the dilatation variance given 
in Eq.~\eqref{thermal_average_energy-momentum_tensor2})
as follows:  
\begin{align} 
&
V_\textrm{s eff} (\varphi) 
\nonumber \\ 
= & V_s(\varphi) + \frac{1}{4} \mathrm{e}^{-4 \sqrt{1/6} \kappa \varphi} 
\nonumber \\ 
& 
\quad \times \sum_in_i\tilde{m}_i^2(\phi)
\left. \left[
\frac{\tilde{M}_i^2(\phi, T)}{16\pi^2}\left(\log \frac{\tilde{M}_i^2(\phi, T)}{\tilde{\mu}^2}-c_i
+\frac{1}{2}\right)
+\frac{T^2}{\pi^2}I'_{B,F}\left(\frac{\tilde{M}_i^2(\phi, T)}{T^2}\right)
\right] \right |_{\phi=v(T)}
\nonumber \\ 
= &
V_s(\varphi) + \frac{1}{4} \mathrm{e}^{-4 \sqrt{1/6} \kappa \varphi} 
\left. \left [
4B\phi^4 \log \frac{\phi^2}{v^2}+
\sum_in_i\tilde{m}_i^2(\phi)
\frac{T^2}{\pi^2}I'_{B,F}\left(\frac{\tilde{M}_i^2(\phi, T)}{T^2} \right)
\right] \right|_{\phi=v(T)}
\label{varphi-h-V:expand0}
\, ,
\end{align}
and 
\begin{align}
& 
V_\textrm{s eff} (\varphi) 
\nonumber \\
\mathop{\simeq}_{\text{high-$T$}} &
V_s(\varphi) + \frac{1}{4} \mathrm{e}^{-4 \sqrt{1/6} \kappa \varphi} 
 \left. \left [
4B\phi^4\ln\frac{\phi^2}{v^2} + \Pi_\phi(T)\phi^2
- 4B\phi^4\ln\frac{\phi^2}{v^2} - 3ET\phi^3 + \lambda_T \phi^4 
+ \cdots 
\right] \right|_{\phi=v(T)} 
\nonumber \\ 
\rightarrow \ & 
V_s(\varphi) + \frac{1}{4} \mathrm{e}^{-4 \sqrt{1/6} \kappa \varphi} 
\left[ 
\Pi_\phi(T) (v(T) + \phi)^2 - 2 E T (v(T) + \phi)^3 + \lambda_T (v(T) + \phi)^4 + \cdots 
\right]
\, . \label{varphi-h-V:expand}
\end{align}    

In arriving at the second equality from the first one in Eq.~\eqref{varphi-h-V:expand0}, 
we have used the tadpole condition $\partial V_{h\, \rm{eff}}/\partial \phi|_{\phi=v(T)} =0$.
In reaching the last expression in Eq.~\eqref{varphi-h-V:expand},
we have curried out the high-temperature expansion (for the unresummed potential),
to make the temperature dependence clearer~\footnote{
\label{foot-I}
The following high-temperature expansions of the thermal functions have been used:  
\begin{align}
I_B(a^{2}) &= 
- \frac{\pi^{4}}{45} + \frac{\pi^2}{12} a^{2} - \frac{\pi}{6} (a^{2})^{\frac{3}{2}}
- \frac{a^{4}}{32} \left( \log \frac{a^{2}}{\alpha_{B}} - \frac{3}{2} \right) + \mathcal{O}(a^{6})
\, , \nonumber \\
I'_B(a^{2}) &= 
\frac{\partial I_{B}(a^{2})}{\partial a^{2}}
= 
\frac{\pi^{2}}{12} - \frac{\pi}{4}(a^{2})^{\frac{1}{2}}
- \frac{a^{2}}{16} \left( \log\frac{a^{2}}{\alpha_B} -1 \right) + \mathcal{O}(a^{4})
\, , \nonumber \\
I_F(a^{2}) &=
\frac{7\pi^{4}}{360} - \frac{\pi^{2}}{24}a^{2}
- \frac{a^{4}}{32} \left( \log \frac{a^{2}}{\alpha_{F}} - \frac{3}{2} \right) + \mathcal{O}(a^{6})
\, , \nonumber \\
I'_F(a^{2}) &=
\frac{\partial I_{F}(a^2)}{\partial a^2}
= - \frac{\pi^{2}}{24} - \frac{a^{2}}{16} \left( \log \frac{a^{2}}{\alpha_{F}} - 1 \right) + \mathcal{O}(a^{4}), 
\nonumber 
\end{align}
where 
$\log \alpha_B = 2 \log (4 \pi) - 2\gamma_E \simeq 3.91$ and 
$\log \alpha_F = 2 \log \pi - 2 \gamma_E \simeq 1.14$. 
}.
We also made the VEV $v(T)$ developed to have the fluctuating Higgs field $\phi$,
and the ellipses denote higher order terms with respect to the $\phi$ field. 
In Eq.~\eqref{varphi-h-V:expand},
the $\Pi_\phi(T)$, $E$ and $\lambda_T$ are defined as 
\begin{align} 
\Pi_\phi (T)& = \sum_{i={\rm bosons}} n_i \frac{m_i^2}{v^2} \frac{T^2}{12} 
- \sum_{i={\rm fermions}} n_i \frac{m_i^2}{v^2} \frac{T^2}{24} 
\,, \nonumber \\ 
E & = \sum_{i={\rm bosons}} n_i \frac{m_i^3}{12 \pi v^3}  
\, , \nonumber \\ 
\lambda_T &= 4 B - \sum_{i} n_i \frac{m_i^4}{16 \pi^2 v^4} \log \frac{m_i^2}{\alpha_{B,F} T^2} 
\, ,
\end{align} 
where the definitions of $\alpha_B$ and $\alpha_F$ being in the footnote~\ref{foot-I}.

Note that the $\phi$-tadpole terms in Eq.(\ref{varphi-h-V:expand}) should be eliminated by imposing 
the tadpole condition $\partial V_{h\, \rm{eff}}/\partial v =0$.
Obviously, the scale symmetry is not respected in the induced scalaron-coupling form, 
which has been spontaneously (radiatively) broken by the Higgs VEV $v$ 
and explicitly broken by the renormalization (i.e. the Higgs VEV) and introduction of temperature~\footnote{
One could absorb those explicit breaking effects by introducing 
the running coupling constants regarding the renormalization scale 
and temperature, into a manifestly renormalized form for the effective potential.  
}.
As expected from Eq.(\ref{varphi-h-V:expand}), 
it implies that the chameleon mechanism will be influenced by the coupling with 
the Higgs potential with those net scale-symmetry breaking-effects encoded, 
as will clearly be seen later soon.

A similar discussion on the chameleon mechanism affected by 
the Higgs potential term has been made in~\cite{Burrage:2018dvt} at zero temperature, 
in which the scalaron plays a role of compensator for the scale invariance in the matter sector (i.e. the standard model) 
and develops its VEV at the tree-level of the standard model
to trigger the scale-symmetry breaking and subsequently breaking the EW symmetry.   
In comparison with the earlier work,  
in the present our case, 
the scale symmetry is spontaneously (radiatively) broken by the matter-sector dynamics (SI-2HDM) at the one-loop level, 
and explicitly broken by introduction of the renormalization scale and temperature, 
which gives the significant effect on the chameleon mechanism at around the EWPT epoch ($T \sim T_C$) 
as we will explicitly demonstrate in the next subsection.
A distinct difference from the earlier work can also be observed in
the target-Higgs potential form of logarithmic type including the finite temperature terms 
in the present scenario. 

Nevertheless, one might simply suspect that 
the chameleon mechanism may not significantly be affected 
as was indicated in~\cite{Blas:2011ac} and \cite{Ferreira:2016wem,Ferreira:2016kxi},
unless the scale-symmetry breaking is supplied by emergence of non-derivative couplings (i.e. potential terms);
for instance, the Higgs portal coupling~\cite{Burrage:2018dvt}. 
Going beyond the tree-level, 
this observation might still be operative even including radiative corrections 
if they are regularized by a scale-invariant dimensional regularization~\cite{Ghilencea:2015mza,Ghilencea:2016ckm}.
The scale-invariant dimensional regularization
could wash out scalaron couplings to the Higgs potential independent of the temperature $T$, 
as displayed in Eq.~\eqref{varphi-h-V:expand}.  
However, in the present our scenario,  
the scalaron would still be left with finite temperature terms as seen in Eq.~\eqref{varphi-h-V:expand}, 
which serve as another explicit-breaking source. 

Note also that such a temperature-dependent part will not be moved away 
in contrast to an artificial renormalization-scale dependence, 
which manifests the fact that the scale symmetry for the matte sector is explicitly broken 
once it is put in the the thermal bath with the characteristic temperature.
Moreover, its breaking effect arises to be seen at the loop level of the thermal field theory. 

\subsection{Comparison with conventional fluid picture}
\label{4C}

\begin{figure}[htbp]
\centering
\includegraphics[width=0.6\textwidth]{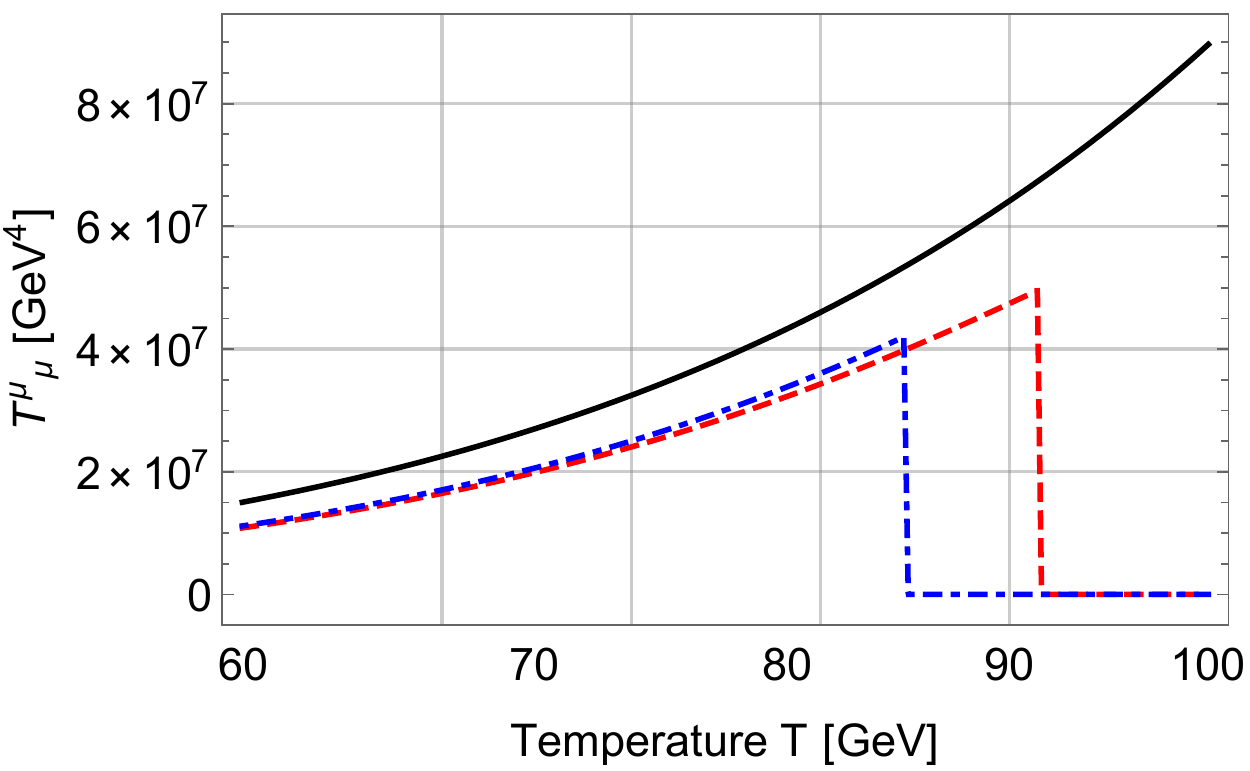}
\caption{
The dashed red and dot-dashed blue lines show the trace of the energy-momentum tensor with (without) resummation prescription, respectively.
The solid black line shows the trace of the energy-momentum tensor constructed only from the standard-model-particles in 
the conventional fluid approach. 
The coincidence in magnitude between the solid black and other curves 
imply the validity of the approximation and the thermal decoupling of heavier 
Higgses with the mass above 100 [GeV]. 
}
\label{trace_energy-momentum_fig}
\end{figure}

Now we evaluate the 
$T^{\mu}_{\ \mu}$
in Eq.(\ref{thermal_average_energy-momentum_tensor})   
in the SI-2HDM based on Eq.~\eqref{thermal_average_energy-momentum_tensor2} with respect to the temperature $T$,  
to obtain the plot in Fig.~\ref{trace_energy-momentum_fig}. 
The $T^{\mu}_{\ \mu}$ has been set to exactly zero before the EWPT
because of the potential convention for the sphaleron 
freeze-out analysis (i.e. $V(\phi, T) \to V(\phi, T) - V(\phi=0, T$) as was done in Sec.III C), 
in which thermally-driven vacuum-energy terms, possibly present in the EW symmetric phase, have been dropped.
That is, the scale-invariant limit has been achieved in the symmetric phase. 
Here, we have applied two different ways (the dashed red and dot-dashed blue lines) 
in taking into account the resummation prescription.
In both two cases, the SI-2HDM model predicts a sharp dump at around $80 - 90 [\mathrm{GeV}]$, which indeed reflects the strong-first order EWPT, 
and the trace of the energy-momentum tensor becomes zero as we expected.
We also plot the result in the previous study \cite{Katsuragawa:2017wge} 
where we calculate the trace of the energy-momentum tensor comprised only 
by the standard model particles in the conventional fluid approximation (the solid black line).

Figure.~\ref{trace_energy-momentum_fig} clearly demonstrates that 
the SI-2HDM with the exact analysis based on the quantum field theory does not show a large deviation 
from the approximated fluid description in the trace of energy-momentum tensor, though they are somewhat different by a factor of order one.
Thus, it has been shown that the conventional approach actually works appropriately
in the evaluation of the chameleon mechanism although it cannot describe the EWPT.

\section{Chameleon mechanism over the EWPT}
\label{5}

In this section, 
we formulate the chameleon mechanism in the environment of 
the EWPT in the early Universe. 
Then, we apply it to the evaluation of the scalaron mass and potential 
in the EWPT environment, 
where the SI-2HDM plays a significant role as we introduced in the previous section.

\subsection{ Model of $F(R)$ gravity}
\label{5A}

In order to examine the chameleon mechanism in the EWPT background,
we consider the following model \cite{Starobinsky:2007hu},
\begin{align}
F(R) = R - \beta R_{c} \left[ 1 - \left( 1 +  \frac{R^{2}}{R^{2}_{c}} \right)^{-n} \right] + \alpha R^{2}
\label{starobinsky_action1}
\, .
\end{align}
The $R_{c}$ is taken to be a typical energy scale,  
where the gravitational action deviates from the Einstein-Hilbert action, 
and one expects $ R_{c} \sim \Lambda \simeq 4 \times 10^{-84} [\mathrm{GeV}^{2}]$.
The index $n$ and the parameter $\beta$ are chosen to be positive constants.  
The $\alpha$ expresses another high energy scale.
It has been known that the $F(R)$ gravity models for the dark energy generally suffer from the curvature singularity problem \cite{Frolov:2008uf}.
This problem can be cured with $R^{2}$ correction \cite{Dev:2008rx,Kobayashi:2008wc} ,
and the scalaron mass is upper-bounded and becomes finite in the high-density region \cite{Thongkool:2009js,Katsuragawa:2017wge}.
Note that $R^{2}$ term is not necessarily identified with the part of $R^{2}$ inflation model.

The curvature scale $R$ should be larger than the dark energy scale $R_{c}$ because the chameleon mechanism works in the presence of matters.
Therefore, we work in the large curvature limit $R_{c}<R$.  
In the large curvature limit, 
the minimum of the potential is determined with Eq.~(\ref{potentialminimum1})
\begin{align}
0
&=
R_{\min} - 2 \beta R_{c} + 2 (n+1) \beta R_{c} \left( \frac{R_{\min}}{R_{c}} \right)^{-2n} + \kappa^{2} T^{\mu}_{\ \mu}
\label{potentialminimum2}
\, .
\end{align}
Note that the $\alpha R^2$ does not affect the stationary condition.
The second and third terms in Eq.~\eqref{potentialminimum2} are negligible in the large curvature limit $R_{c}<R_{\min}$, 
and one finds
\begin{align}
R_{\min} \approx - \kappa^{2} T^{\mu}_{\ \mu}
\, .
\end{align}

As an illustration, we assume that the matter contribution is approximately expressed as the pressure-less dust, 
$T^{\mu}_{\ \mu}=-\rho$, where $\rho$ is the matter energy density.
Then, the scalaron mass in Eq.~(\ref{mass:formula1}) 
is evaluated in the large curvature limit as 
\begin{align}
m_{\varphi}^{2} (\rho)
& \approx
\frac{R_{c}}{6n (2n+1) \beta} \left[ 
\left( \frac{\kappa^{2} \rho}{R_{c} }  \right)^{-2(n+1)}
+ \frac{\alpha R_{c}}{n(2n+1)\beta} 
\right]^{-1} 
\frac{1}{1 + 2 \alpha \kappa^{2} \rho}
\label{scalaronmass1}
\, .
\end{align}
We find that the scalaron mass is given by the increasing function of the energy density $\rho$, 
and thus, the scalaron becomes heavy in the high-density region of matter as we expected.
In the following analysis,
we use this $F(R)$ model to study the chameleon mechanism and effective potential of the scalaron
in the EWPT environment.

\subsection{Chameleon mechanism in the EWPT environment}
\label{5B}

\begin{figure}[htbp]
\centering
\includegraphics[width=0.6\textwidth]{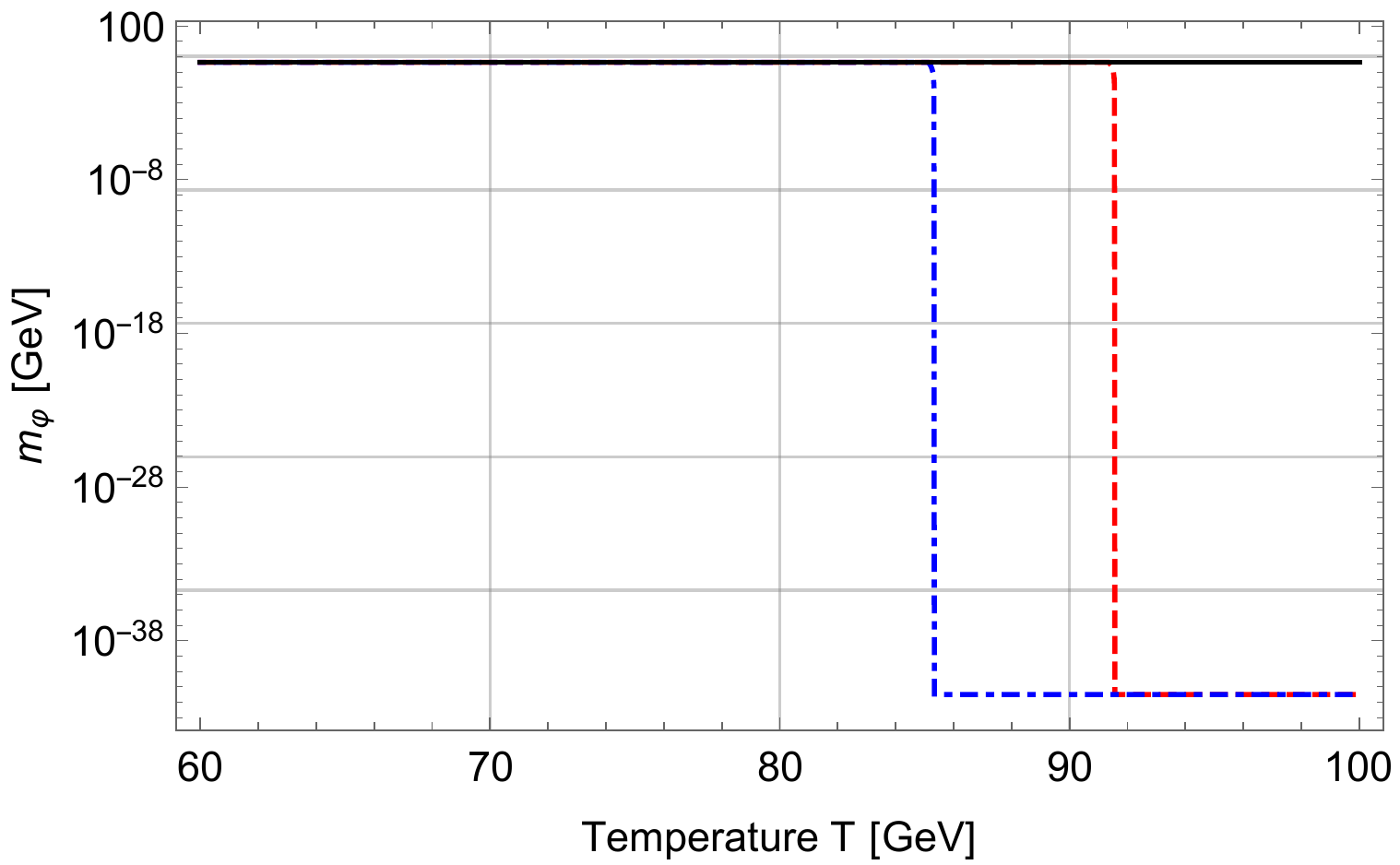}
\caption{
The dashed red and dot-dashed blue lines show the scalaron mass for SI-2HDM with (without) resummation prescription, respectively.
The solid black line shows the scalaron mass in the conventional fluid approach.
The parameters of the $F(R)$ function are chosen as $n=1$, $\beta=2$, 
$\alpha=1[\mathrm{GeV}^{-2}]$.  
}
\label{scalaron_mass_fig}
\end{figure}

Next, we convert the trace of energy-momentum tensor as in Fig. \ref{trace_energy-momentum_fig}
into the temperature-dependence of the scalaron mass.
We show the result in Fig.~\ref{scalaron_mass_fig}.
As an illustration, we take the parameters in Eq.~\eqref{starobinsky_action1}
as $n=1$, $\beta=2$, $\alpha=1[\mathrm{GeV}^{-2}]$.
Since the trace of energy-momentum tensor is zero before the EWPT,
the dashed red and dot-dashed blue lines show that
the scalaron mass is given by the dark energy scale,
$m_{\varphi} \sim 3\times10^{-33}[\mathrm{eV}]$.
After the EWPT,
the effective potential of scalaron achieves the finite trace of energy-momentum tensor,
and the chameleon mechanism makes the scalaron mass heavy.
The scalaron mass after the EWPT takes the constant value as in \cite{Katsuragawa:2017wge}, 
$m_{\varphi} \sim 0.1[\mathrm{GeV}]$,
for the choice of $\alpha=1[\mathrm{GeV}^{-2}]$.

We note that
the constancy of the scalaron mass is due to $R^{2}$ term in the $F(R)$ model Eq.~\eqref{starobinsky_action1}.
When we work in a large curvature limit where $R_{c}<R<1/\alpha$,
we find that the mass formula Eq.~\eqref{scalaronmass1} is reduced to
\begin{align}
m^{2}_{\varphi} \approx \frac{1}{6\alpha}
\, .
\end{align}
This is why the scalaron mass $m_{\varphi}$ is approximately computed to be  $\sim  0.1[\mathrm{GeV}]$ when $\alpha=1[\mathrm{GeV}^{-2}]$.
It is also noted that
the scalaron mass becomes a trans-Planckian scale without $R^{2}$ corrections, which is related to the singularity problem in the $F(R)$ gravity.

\subsection{Scalaron potential in the EWPT environment}
\label{5C}

\begin{figure}[htbp]
\centering
\includegraphics[width=0.6\textwidth]{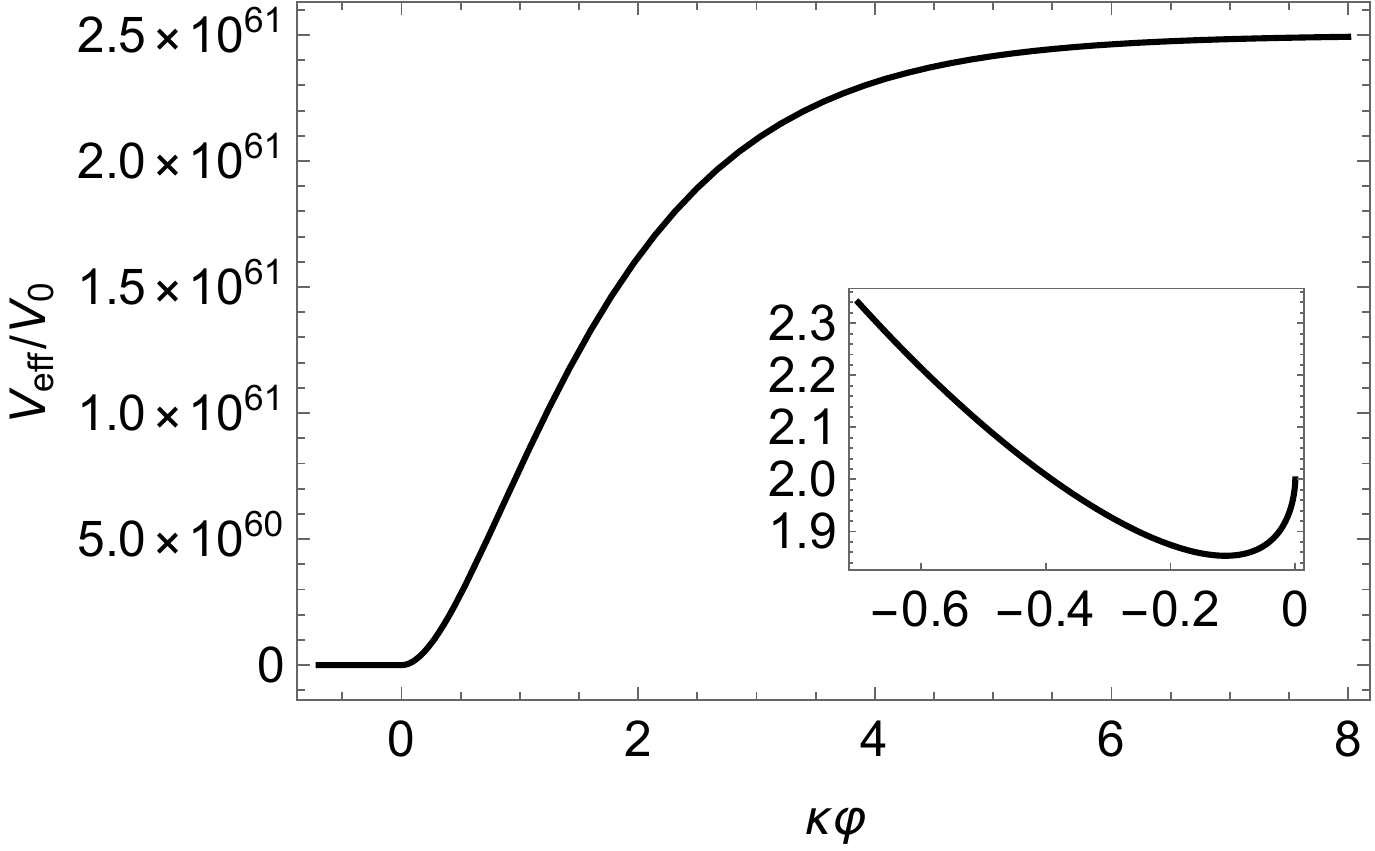}
\caption{
The solid black line shows the effective potential of the scalar field
with $n=1$, $\beta=2$, and $\alpha = 10^{22}[\mathrm{GeV}^{-2}]$.
The potential is normalized by $V_{0} = \frac{R_{c}}{2\kappa^{2}} \sim \rho_{\Lambda}$
where $\rho_{\Lambda}$ is the dark energy density.
Before the EWPT, the SI-2HDM does not generate the trace of energy-momentum tensor,
and thus, the chameleon mechanism does not work.
}
\label{potential_fig1}
\end{figure}

\begin{figure}[htbp]
\centering
\includegraphics[width=0.6\textwidth]{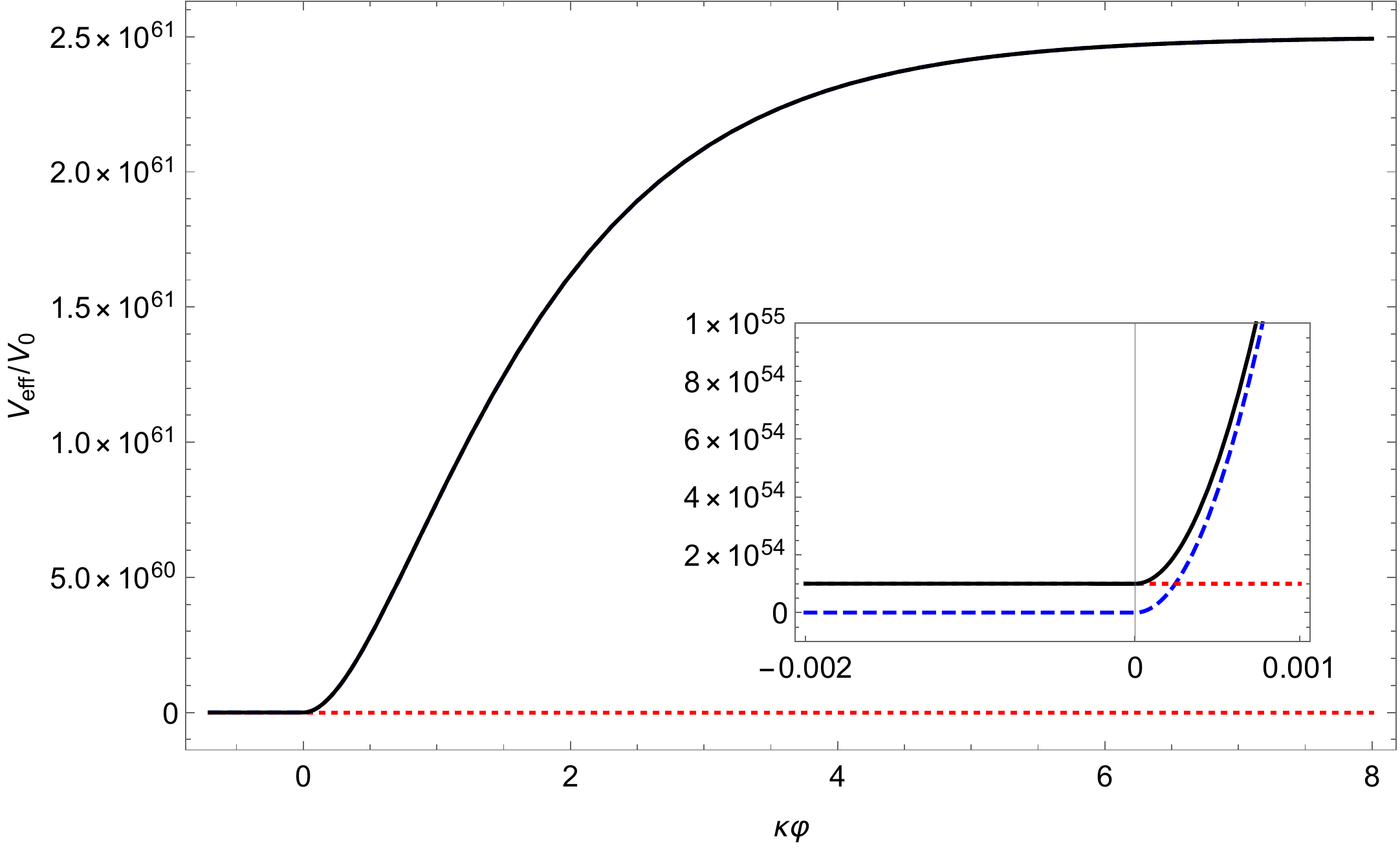}
\caption{
The same parameter choice as in Fig.~\ref{potential_fig1}.
The solid black line shows the effective potential
although the blue dashed line represents the original potential of the scalaron field.
The red dotted line shows the matter contribution.
After the EWPT, the trace of energy-momentum tensor takes the non-zero value,
and affects the effective potential due to the chameleon mechanism.
Immediate the EWPT, $T^{\mu}_{\ \mu} \sim 4 \times 10^{7} [\mathrm{GeV}^{4}]$.
}
\label{potential_fig2}
\end{figure}

Finally, we discuss the scalaron potential over the EWPT environment created 
from the SI-2HDM. 
We consider the effective potential of the scalaron field before and after the EWPT.
Right after the EWPT ($T \lesssim T_C \simeq 91.5$ [GeV]) with and without the resummation prescription,
the trace of energy-momentum tensor keeps almost a constant value 
(See Fig.~\ref{trace_energy-momentum_fig} ).
Hence, by inputting $T^{\mu}_{\ \mu} =0 [\mathrm{GeV}^{4}]$
and $T^{\mu}_{\ \mu} \sim 4 \times 10^{7} [\mathrm{GeV}^{4}]$, 
we plot the form of the effective potential before and after the EWPT,
given in Figs.~\ref{potential_fig1} and \ref{potential_fig2}.

In this analysis, we set $\alpha = 10^{22}[\mathrm{GeV}^{-2}]$,
which corresponds to the experimental upper-bound from the fifth forth experiment~\cite{Cembranos:2008gj}.
Before the EWPT ($T \gtrsim T_C$),
the effective potential does not receive the effect of the chameleon mechanism
because the trace of energy-momentum tensor vanishes.
Then, the potential minimum locates at around $\kappa \varphi \sim - 0.1$.
Immediately after the EWPT ($T \lesssim T_C$),
the chameleon mechanism starts to work due to the non-zero trace of energy-momentum tensor induced by the radiative breaking of the scale and EW symmetries, 
and the effective potential is lifted by the SI-2HDM-matter contributions. 
Then, the potential minimum locates at around $\kappa \varphi \sim 0$. 

From the potential analysis, 
we can conjecture the new thermal history of the scalaron field 
with taking into account the EWPT;
the scalaron field at the original potential minimum
is pushed away from the minimum by the EWPT via the chameleon mechanism,
and the scalaron field would locate around the new potential minimum.
As the Universe expands,
the matter effect is decreasing in the effective potential,
and the potential form is approaching to the original one.
The above scenario gives the nonperturbative effect to the time-evolution of the scalaron field,
and the behavior of the scalaron field after the EWPT is nontrivial,
but expected to be around the potential minimum.


\section{Conclusion and Discussion}
\label{summary}

We have investigated the chameleon mechanism in the early Universe, especially the EWPT epoch, 
with the formulation based on the quantum field theory.
We have utilized a class of general SI-2HDM to describe the EWPT 
and evaluated the one-loop effective potential as a function of a constant background Higgs field.
By interpolating the numerical results,
we have confirmed that 
the trace of energy-momentum tensor in the SI-2HDM shows almost the same temperature-dependence as that in the conventional fluid approach after the EWPT,
as far as the order of magnitude is concerned.
Moreover, we have converted the temperature-dependence of the trace of the energy-momentum tensor into that of the scalaron mass,
to find that the scalaron has tiny mass comparable to the dark energy scale, $m_{\varphi} \sim 10^{-33} [\mathrm{eV}]$,
due to the absence of the chameleon mechanism before the EWPT.
This is reflected by the drastic change of the scalaron potential 
caused by the suddenly increased trace of energy-momentum tensor
as we have shown in the shape of the scalaron potential before and after the EWPT.

A couple of comments and discussions on future prospects 
regarding what we have clarified in the present 
paper are as follows.  
Two of authors have studied the scenario that the scalaron can be dark matter \cite{Katsuragawa:2016yir}.
When we quantize the excitation or perturbation around the potential minimum of the scalaron field, we have the particle picture of the scalaron field.
This scenario relies on the assumption that the scalaron field keeps the harmonic oscillation to act as a dust in the cosmic history,
and thus, the initial condition for the oscillation seemed to be given by hand
in analogy with the harmonic oscillation of the axion dark matter.

Regarding the origin of the harmonic oscillation, 
we could find the species of the oscillation in the present paper.
For the scalaron, the matter sector works as an external field,
and the scalaron field would receive the nonlinear effect to start the forced oscillation
when the external field suddenly changes.
Because the EWPT in the matter sector ``kicks" the scalaron field through the chameleon mechanism, 
we can expect that such a kick generates the harmonic oscillation of the scalaron field. 
We also expect that 
the initial condition for the scalaron field at the original potential minimum
would be wiped out by the matter effect,
to avoid the fine-tuning for the harmonic oscillation of the scalaron.

The kick solution had been already argued in \cite{Brax:2004qh,Erickcek:2013oma}.
This is induced by the comparison to the Hubble friction term in the cosmological background, which causes the continuous change of potential form.
On the other hand, the kick by the EWPT is not the continuous one, and thus, 
we can conclude that we have found another possibility of the kick solution in the early Universe.
We also emphasize that the PT does not happen in the scalaron sector
because the scalaron field decouples from the thermal equilibrium due to the Planck-suppressed couplings to the other fields. 
Thus, the discrete change of the potential shape in the chameleon sector should be regarded as a transition-like phenomena induced by the genuine EWPT in the matter sector.

In relation to the oscillation of the scalaron,
we might find the intriguing effect in the phenomenology.
As we have mentioned,
the EWPT deforms the potential shape of the scalaron field,
which may generate the oscillation of the scalaron field.
Since the scalaron field originates from the gravity sector through the Weyl transformation,
the oscillation of the scalaron field would imply generation of an intrinsic gravitational wave. 
It has been suggested that the $F(R)$ gravity predicts the scalar mode of the gravitational waves \cite{Corda:2008jt,Capozziello:2008rq}, 
which is described by the fluctuation of the scalaron field around the potential minimum. 
Therefore, we may expect the origin of the non-tensorial polarization of the gravitational waves induced by the EWPT.
Besides the well-known fact that a strong-first order PT can directly generate the gravitational waves~\cite{Kosowsky:1991ua,Kosowsky:1992rz,Kosowsky:1992vn,Kamionkowski:1993fg}, 
our analysis gives another insight in that the EWPT could indirectly generate the scalar mode of the gravitationalt waves in the $F(R)$ gravity. 
This speculation would open the brand-new phenomenology to explore the beyond-standard-physics in both particle and gravitational physics.


\begin{acknowledgments}
T.K. is supported by International Postdoctoral Exchange Fellowship Program at Central China Normal University
and Project funded by China Postdoctoral Science Foundation 2018M632895. 
S.M. is supported in part by the JSPS Grant-in-Aid for Young Scientists (B) No. 15K17645,
National Science Foundation of China (NSFC) under Grant No. 11747308, and the Seeds Funding of Jilin University.
E.S. is supported by IBS under the project code, IBS-R018-D1.
\end{acknowledgments}

\bibliography{FREWPT-bib}

\begin{thebibliography}{55}%
\makeatletter
\providecommand \@ifxundefined [1]{%
 \@ifx{#1\undefined}
}%
\providecommand \@ifnum [1]{%
 \ifnum #1\expandafter \@firstoftwo
 \else \expandafter \@secondoftwo
 \fi
}%
\providecommand \@ifx [1]{%
 \ifx #1\expandafter \@firstoftwo
 \else \expandafter \@secondoftwo
 \fi
}%
\providecommand \natexlab [1]{#1}%
\providecommand \enquote  [1]{``#1''}%
\providecommand \bibnamefont  [1]{#1}%
\providecommand \bibfnamefont [1]{#1}%
\providecommand \citenamefont [1]{#1}%
\providecommand \href@noop [0]{\@secondoftwo}%
\providecommand \href [0]{\begingroup \@sanitize@url \@href}%
\providecommand \@href[1]{\@@startlink{#1}\@@href}%
\providecommand \@@href[1]{\endgroup#1\@@endlink}%
\providecommand \@sanitize@url [0]{\catcode `\\12\catcode `\$12\catcode
  `\&12\catcode `\#12\catcode `\^12\catcode `\_12\catcode `\%12\relax}%
\providecommand \@@startlink[1]{}%
\providecommand \@@endlink[0]{}%
\providecommand \url  [0]{\begingroup\@sanitize@url \@url }%
\providecommand \@url [1]{\endgroup\@href {#1}{\urlprefix }}%
\providecommand \urlprefix  [0]{URL }%
\providecommand \Eprint [0]{\href }%
\providecommand \doibase [0]{http://dx.doi.org/}%
\providecommand \selectlanguage [0]{\@gobble}%
\providecommand \bibinfo  [0]{\@secondoftwo}%
\providecommand \bibfield  [0]{\@secondoftwo}%
\providecommand \translation [1]{[#1]}%
\providecommand \BibitemOpen [0]{}%
\providecommand \bibitemStop [0]{}%
\providecommand \bibitemNoStop [0]{.\EOS\space}%
\providecommand \EOS [0]{\spacefactor3000\relax}%
\providecommand \BibitemShut  [1]{\csname bibitem#1\endcsname}%
\let\auto@bib@innerbib\@empty
\bibitem [{\citenamefont {Clifton}\ \emph {et~al.}(2012)\citenamefont
  {Clifton}, \citenamefont {Ferreira}, \citenamefont {Padilla},\ and\
  \citenamefont {Skordis}}]{Clifton:2011jh}%
  \BibitemOpen
  \bibfield  {author} {\bibinfo {author} {\bibfnamefont {T.}~\bibnamefont
  {Clifton}}, \bibinfo {author} {\bibfnamefont {P.~G.}\ \bibnamefont
  {Ferreira}}, \bibinfo {author} {\bibfnamefont {A.}~\bibnamefont {Padilla}}, \
  and\ \bibinfo {author} {\bibfnamefont {C.}~\bibnamefont {Skordis}},\ }\href
  {\doibase 10.1016/j.physrep.2012.01.001} {\bibfield  {journal} {\bibinfo
  {journal} {Phys. Rept.}\ }\textbf {\bibinfo {volume} {513}},\ \bibinfo
  {pages} {1} (\bibinfo {year} {2012})},\ \Eprint
  {http://arxiv.org/abs/1106.2476} {arXiv:1106.2476 [astro-ph.CO]} \BibitemShut
  {NoStop}%
\bibitem [{\citenamefont {Nojiri}\ and\ \citenamefont
  {Odintsov}(2011)}]{Nojiri:2010wj}%
  \BibitemOpen
  \bibfield  {author} {\bibinfo {author} {\bibfnamefont {S.}~\bibnamefont
  {Nojiri}}\ and\ \bibinfo {author} {\bibfnamefont {S.~D.}\ \bibnamefont
  {Odintsov}},\ }\href {\doibase 10.1016/j.physrep.2011.04.001} {\bibfield
  {journal} {\bibinfo  {journal} {Phys. Rept.}\ }\textbf {\bibinfo {volume}
  {505}},\ \bibinfo {pages} {59} (\bibinfo {year} {2011})},\ \Eprint
  {http://arxiv.org/abs/1011.0544} {arXiv:1011.0544 [gr-qc]} \BibitemShut
  {NoStop}%
\bibitem [{\citenamefont {Joyce}\ \emph {et~al.}(2015)\citenamefont {Joyce},
  \citenamefont {Jain}, \citenamefont {Khoury},\ and\ \citenamefont
  {Trodden}}]{Joyce:2014kja}%
  \BibitemOpen
  \bibfield  {author} {\bibinfo {author} {\bibfnamefont {A.}~\bibnamefont
  {Joyce}}, \bibinfo {author} {\bibfnamefont {B.}~\bibnamefont {Jain}},
  \bibinfo {author} {\bibfnamefont {J.}~\bibnamefont {Khoury}}, \ and\ \bibinfo
  {author} {\bibfnamefont {M.}~\bibnamefont {Trodden}},\ }\href {\doibase
  10.1016/j.physrep.2014.12.002} {\bibfield  {journal} {\bibinfo  {journal}
  {Phys. Rept.}\ }\textbf {\bibinfo {volume} {568}},\ \bibinfo {pages} {1}
  (\bibinfo {year} {2015})},\ \Eprint {http://arxiv.org/abs/1407.0059}
  {arXiv:1407.0059 [astro-ph.CO]} \BibitemShut {NoStop}%
\bibitem [{\citenamefont {Berti}\ \emph {et~al.}(2015)\citenamefont {Berti}
  \emph {et~al.}}]{Berti:2015itd}%
  \BibitemOpen
  \bibfield  {author} {\bibinfo {author} {\bibfnamefont {E.}~\bibnamefont
  {Berti}} \emph {et~al.},\ }\href {\doibase 10.1088/0264-9381/32/24/243001}
  {\bibfield  {journal} {\bibinfo  {journal} {Class. Quant. Grav.}\ }\textbf
  {\bibinfo {volume} {32}},\ \bibinfo {pages} {243001} (\bibinfo {year}
  {2015})},\ \Eprint {http://arxiv.org/abs/1501.07274} {arXiv:1501.07274
  [gr-qc]} \BibitemShut {NoStop}%
\bibitem [{\citenamefont {Koyama}(2016)}]{Koyama:2015vza}%
  \BibitemOpen
  \bibfield  {author} {\bibinfo {author} {\bibfnamefont {K.}~\bibnamefont
  {Koyama}},\ }\href {\doibase 10.1088/0034-4885/79/4/046902} {\bibfield
  {journal} {\bibinfo  {journal} {Rept. Prog. Phys.}\ }\textbf {\bibinfo
  {volume} {79}},\ \bibinfo {pages} {046902} (\bibinfo {year} {2016})},\
  \Eprint {http://arxiv.org/abs/1504.04623} {arXiv:1504.04623 [astro-ph.CO]}
  \BibitemShut {NoStop}%
\bibitem [{\citenamefont {Bull}\ \emph {et~al.}(2016)\citenamefont {Bull} \emph
  {et~al.}}]{Bull:2015stt}%
  \BibitemOpen
  \bibfield  {author} {\bibinfo {author} {\bibfnamefont {P.}~\bibnamefont
  {Bull}} \emph {et~al.},\ }\href {\doibase 10.1016/j.dark.2016.02.001}
  {\bibfield  {journal} {\bibinfo  {journal} {Phys. Dark Univ.}\ }\textbf
  {\bibinfo {volume} {12}},\ \bibinfo {pages} {56} (\bibinfo {year} {2016})},\
  \Eprint {http://arxiv.org/abs/1512.05356} {arXiv:1512.05356 [astro-ph.CO]}
  \BibitemShut {NoStop}%
\bibitem [{\citenamefont {Brax}(2013)}]{Brax:2013ida}%
  \BibitemOpen
  \bibfield  {author} {\bibinfo {author} {\bibfnamefont {P.}~\bibnamefont
  {Brax}},\ }\href {\doibase 10.1088/0264-9381/30/21/214005} {\bibfield
  {journal} {\bibinfo  {journal} {Class. Quant. Grav.}\ }\textbf {\bibinfo
  {volume} {30}},\ \bibinfo {pages} {214005} (\bibinfo {year}
  {2013})}\BibitemShut {NoStop}%
\bibitem [{\citenamefont {Burrage}\ and\ \citenamefont
  {Sakstein}(2018)}]{Burrage:2017qrf}%
  \BibitemOpen
  \bibfield  {author} {\bibinfo {author} {\bibfnamefont {C.}~\bibnamefont
  {Burrage}}\ and\ \bibinfo {author} {\bibfnamefont {J.}~\bibnamefont
  {Sakstein}},\ }\href {\doibase 10.1007/s41114-018-0011-x} {\bibfield
  {journal} {\bibinfo  {journal} {Living Rev. Rel.}\ }\textbf {\bibinfo
  {volume} {21}},\ \bibinfo {pages} {1} (\bibinfo {year} {2018})},\ \Eprint
  {http://arxiv.org/abs/1709.09071} {arXiv:1709.09071 [astro-ph.CO]}
  \BibitemShut {NoStop}%
\bibitem [{\citenamefont {Khoury}\ and\ \citenamefont
  {Weltman}(2004{\natexlab{a}})}]{Khoury:2003aq}%
  \BibitemOpen
  \bibfield  {author} {\bibinfo {author} {\bibfnamefont {J.}~\bibnamefont
  {Khoury}}\ and\ \bibinfo {author} {\bibfnamefont {A.}~\bibnamefont
  {Weltman}},\ }\href {\doibase 10.1103/PhysRevLett.93.171104} {\bibfield
  {journal} {\bibinfo  {journal} {Phys. Rev. Lett.}\ }\textbf {\bibinfo
  {volume} {93}},\ \bibinfo {pages} {171104} (\bibinfo {year}
  {2004}{\natexlab{a}})},\ \Eprint {http://arxiv.org/abs/astro-ph/0309300}
  {arXiv:astro-ph/0309300 [astro-ph]} \BibitemShut {NoStop}%
\bibitem [{\citenamefont {Khoury}\ and\ \citenamefont
  {Weltman}(2004{\natexlab{b}})}]{Khoury:2003rn}%
  \BibitemOpen
  \bibfield  {author} {\bibinfo {author} {\bibfnamefont {J.}~\bibnamefont
  {Khoury}}\ and\ \bibinfo {author} {\bibfnamefont {A.}~\bibnamefont
  {Weltman}},\ }\href {\doibase 10.1103/PhysRevD.69.044026} {\bibfield
  {journal} {\bibinfo  {journal} {Phys. Rev.}\ }\textbf {\bibinfo {volume}
  {D69}},\ \bibinfo {pages} {044026} (\bibinfo {year} {2004}{\natexlab{b}})},\
  \Eprint {http://arxiv.org/abs/astro-ph/0309411} {arXiv:astro-ph/0309411
  [astro-ph]} \BibitemShut {NoStop}%
\bibitem [{\citenamefont {Katsuragawa}\ and\ \citenamefont
  {Matsuzaki}(2018)}]{Katsuragawa:2017wge}%
  \BibitemOpen
  \bibfield  {author} {\bibinfo {author} {\bibfnamefont {T.}~\bibnamefont
  {Katsuragawa}}\ and\ \bibinfo {author} {\bibfnamefont {S.}~\bibnamefont
  {Matsuzaki}},\ }\href {\doibase 10.1103/PhysRevD.97.129902,
  10.1103/PhysRevD.97.064037} {\bibfield  {journal} {\bibinfo  {journal} {Phys.
  Rev.}\ }\textbf {\bibinfo {volume} {D97}},\ \bibinfo {pages} {064037}
  (\bibinfo {year} {2018})},\ \bibinfo {note} {[Erratum: Phys.
  Rev.D97,no.12,129902(2018)]},\ \Eprint {http://arxiv.org/abs/1708.08702}
  {arXiv:1708.08702 [gr-qc]} \BibitemShut {NoStop}%
\bibitem [{\citenamefont {Blas}\ \emph {et~al.}(2011)\citenamefont {Blas},
  \citenamefont {Shaposhnikov},\ and\ \citenamefont
  {Zenhausern}}]{Blas:2011ac}%
  \BibitemOpen
  \bibfield  {author} {\bibinfo {author} {\bibfnamefont {D.}~\bibnamefont
  {Blas}}, \bibinfo {author} {\bibfnamefont {M.}~\bibnamefont {Shaposhnikov}},
  \ and\ \bibinfo {author} {\bibfnamefont {D.}~\bibnamefont {Zenhausern}},\
  }\href {\doibase 10.1103/PhysRevD.84.044001} {\bibfield  {journal} {\bibinfo
  {journal} {Phys. Rev.}\ }\textbf {\bibinfo {volume} {D84}},\ \bibinfo {pages}
  {044001} (\bibinfo {year} {2011})},\ \Eprint {http://arxiv.org/abs/1104.1392}
  {arXiv:1104.1392 [hep-th]} \BibitemShut {NoStop}%
\bibitem [{\citenamefont {Ferreira}\ \emph
  {et~al.}(2017{\natexlab{a}})\citenamefont {Ferreira}, \citenamefont {Hill},\
  and\ \citenamefont {Ross}}]{Ferreira:2016wem}%
  \BibitemOpen
  \bibfield  {author} {\bibinfo {author} {\bibfnamefont {P.~G.}\ \bibnamefont
  {Ferreira}}, \bibinfo {author} {\bibfnamefont {C.~T.}\ \bibnamefont {Hill}},
  \ and\ \bibinfo {author} {\bibfnamefont {G.~G.}\ \bibnamefont {Ross}},\
  }\href {\doibase 10.1103/PhysRevD.95.043507} {\bibfield  {journal} {\bibinfo
  {journal} {Phys. Rev.}\ }\textbf {\bibinfo {volume} {D95}},\ \bibinfo {pages}
  {043507} (\bibinfo {year} {2017}{\natexlab{a}})},\ \Eprint
  {http://arxiv.org/abs/1610.09243} {arXiv:1610.09243 [hep-th]} \BibitemShut
  {NoStop}%
\bibitem [{\citenamefont {Ferreira}\ \emph
  {et~al.}(2017{\natexlab{b}})\citenamefont {Ferreira}, \citenamefont {Hill},\
  and\ \citenamefont {Ross}}]{Ferreira:2016kxi}%
  \BibitemOpen
  \bibfield  {author} {\bibinfo {author} {\bibfnamefont {P.~G.}\ \bibnamefont
  {Ferreira}}, \bibinfo {author} {\bibfnamefont {C.~T.}\ \bibnamefont {Hill}},
  \ and\ \bibinfo {author} {\bibfnamefont {G.~G.}\ \bibnamefont {Ross}},\
  }\href {\doibase 10.1103/PhysRevD.95.064038} {\bibfield  {journal} {\bibinfo
  {journal} {Phys. Rev.}\ }\textbf {\bibinfo {volume} {D95}},\ \bibinfo {pages}
  {064038} (\bibinfo {year} {2017}{\natexlab{b}})},\ \Eprint
  {http://arxiv.org/abs/1612.03157} {arXiv:1612.03157 [gr-qc]} \BibitemShut
  {NoStop}%
\bibitem [{\citenamefont {Ghilencea}(2016)}]{Ghilencea:2015mza}%
  \BibitemOpen
  \bibfield  {author} {\bibinfo {author} {\bibfnamefont {D.~M.}\ \bibnamefont
  {Ghilencea}},\ }\href {\doibase 10.1103/PhysRevD.93.105006} {\bibfield
  {journal} {\bibinfo  {journal} {Phys. Rev.}\ }\textbf {\bibinfo {volume}
  {D93}},\ \bibinfo {pages} {105006} (\bibinfo {year} {2016})},\ \Eprint
  {http://arxiv.org/abs/1508.00595} {arXiv:1508.00595 [hep-ph]} \BibitemShut
  {NoStop}%
\bibitem [{\citenamefont {Ghilencea}\ \emph {et~al.}(2016)\citenamefont
  {Ghilencea}, \citenamefont {Lalak},\ and\ \citenamefont
  {Olszewski}}]{Ghilencea:2016ckm}%
  \BibitemOpen
  \bibfield  {author} {\bibinfo {author} {\bibfnamefont {D.~M.}\ \bibnamefont
  {Ghilencea}}, \bibinfo {author} {\bibfnamefont {Z.}~\bibnamefont {Lalak}}, \
  and\ \bibinfo {author} {\bibfnamefont {P.}~\bibnamefont {Olszewski}},\ }\href
  {\doibase 10.1140/epjc/s10052-016-4475-0} {\bibfield  {journal} {\bibinfo
  {journal} {Eur. Phys. J.}\ }\textbf {\bibinfo {volume} {C76}},\ \bibinfo
  {pages} {656} (\bibinfo {year} {2016})},\ \Eprint
  {http://arxiv.org/abs/1608.05336} {arXiv:1608.05336 [hep-th]} \BibitemShut
  {NoStop}%
\bibitem [{\citenamefont {Fuyuto}\ and\ \citenamefont
  {Senaha}(2015)}]{Fuyuto:2015vna}%
  \BibitemOpen
  \bibfield  {author} {\bibinfo {author} {\bibfnamefont {K.}~\bibnamefont
  {Fuyuto}}\ and\ \bibinfo {author} {\bibfnamefont {E.}~\bibnamefont
  {Senaha}},\ }\href {\doibase 10.1016/j.physletb.2015.05.061} {\bibfield
  {journal} {\bibinfo  {journal} {Phys. Lett.}\ }\textbf {\bibinfo {volume}
  {B747}},\ \bibinfo {pages} {152} (\bibinfo {year} {2015})}\BibitemShut
  {NoStop}%
\bibitem [{\citenamefont {Fuyuto}\ \emph {et~al.}(2018)\citenamefont {Fuyuto},
  \citenamefont {Hou},\ and\ \citenamefont {Senaha}}]{Fuyuto:2017ewj}%
  \BibitemOpen
  \bibfield  {author} {\bibinfo {author} {\bibfnamefont {K.}~\bibnamefont
  {Fuyuto}}, \bibinfo {author} {\bibfnamefont {W.-S.}\ \bibnamefont {Hou}}, \
  and\ \bibinfo {author} {\bibfnamefont {E.}~\bibnamefont {Senaha}},\ }\href
  {\doibase 10.1016/j.physletb.2017.11.073} {\bibfield  {journal} {\bibinfo
  {journal} {Phys. Lett.}\ }\textbf {\bibinfo {volume} {B776}},\ \bibinfo
  {pages} {402} (\bibinfo {year} {2018})},\ \Eprint
  {http://arxiv.org/abs/1705.05034} {arXiv:1705.05034 [hep-ph]} \BibitemShut
  {NoStop}%
\bibitem [{\citenamefont {Modak}\ and\ \citenamefont
  {Senaha}(2018)}]{Modak:2018csw}%
  \BibitemOpen
  \bibfield  {author} {\bibinfo {author} {\bibfnamefont {T.}~\bibnamefont
  {Modak}}\ and\ \bibinfo {author} {\bibfnamefont {E.}~\bibnamefont {Senaha}},\
  }\href@noop {} {\  (\bibinfo {year} {2018})},\ \Eprint
  {http://arxiv.org/abs/1811.08088} {arXiv:1811.08088 [hep-ph]} \BibitemShut
  {NoStop}%
\bibitem [{\citenamefont {Kuzmin}\ \emph {et~al.}(1985)\citenamefont {Kuzmin},
  \citenamefont {Rubakov},\ and\ \citenamefont {Shaposhnikov}}]{Kuzmin:1985mm}%
  \BibitemOpen
  \bibfield  {author} {\bibinfo {author} {\bibfnamefont {V.~A.}\ \bibnamefont
  {Kuzmin}}, \bibinfo {author} {\bibfnamefont {V.~A.}\ \bibnamefont {Rubakov}},
  \ and\ \bibinfo {author} {\bibfnamefont {M.~E.}\ \bibnamefont
  {Shaposhnikov}},\ }\href {\doibase 10.1016/0370-2693(85)91028-7} {\bibfield
  {journal} {\bibinfo  {journal} {Phys. Lett.}\ }\textbf {\bibinfo {volume}
  {155B}},\ \bibinfo {pages} {36} (\bibinfo {year} {1985})}\BibitemShut
  {NoStop}%
\bibitem [{\citenamefont {Quiros}(1994)}]{Quiros:1994dr}%
  \BibitemOpen
  \bibfield  {author} {\bibinfo {author} {\bibfnamefont {M.}~\bibnamefont
  {Quiros}},\ }\href@noop {} {\bibfield  {journal} {\bibinfo  {journal} {Helv.
  Phys. Acta}\ }\textbf {\bibinfo {volume} {67}},\ \bibinfo {pages} {451}
  (\bibinfo {year} {1994})}\BibitemShut {NoStop}%
\bibitem [{\citenamefont {Rubakov}\ and\ \citenamefont
  {Shaposhnikov}(1996)}]{Rubakov:1996vz}%
  \BibitemOpen
  \bibfield  {author} {\bibinfo {author} {\bibfnamefont {V.~A.}\ \bibnamefont
  {Rubakov}}\ and\ \bibinfo {author} {\bibfnamefont {M.~E.}\ \bibnamefont
  {Shaposhnikov}},\ }\href {\doibase 10.1070/PU1996v039n05ABEH000145}
  {\bibfield  {journal} {\bibinfo  {journal} {Usp. Fiz. Nauk}\ }\textbf
  {\bibinfo {volume} {166}},\ \bibinfo {pages} {493} (\bibinfo {year}
  {1996})},\ \bibinfo {note} {[Phys. Usp.39,461(1996)]},\ \Eprint
  {http://arxiv.org/abs/hep-ph/9603208} {arXiv:hep-ph/9603208 [hep-ph]}
  \BibitemShut {NoStop}%
\bibitem [{\citenamefont {Funakubo}(1996)}]{Funakubo:1996dw}%
  \BibitemOpen
  \bibfield  {author} {\bibinfo {author} {\bibfnamefont {K.}~\bibnamefont
  {Funakubo}},\ }\href {\doibase 10.1143/PTP.96.475} {\bibfield  {journal}
  {\bibinfo  {journal} {Prog. Theor. Phys.}\ }\textbf {\bibinfo {volume}
  {96}},\ \bibinfo {pages} {475} (\bibinfo {year} {1996})},\ \Eprint
  {http://arxiv.org/abs/hep-ph/9608358} {arXiv:hep-ph/9608358 [hep-ph]}
  \BibitemShut {NoStop}%
\bibitem [{\citenamefont {Morrissey}\ and\ \citenamefont
  {Ramsey-Musolf}(2012)}]{Morrissey:2012db}%
  \BibitemOpen
  \bibfield  {author} {\bibinfo {author} {\bibfnamefont {D.~E.}\ \bibnamefont
  {Morrissey}}\ and\ \bibinfo {author} {\bibfnamefont {M.~J.}\ \bibnamefont
  {Ramsey-Musolf}},\ }\href {\doibase 10.1088/1367-2630/14/12/125003}
  {\bibfield  {journal} {\bibinfo  {journal} {New J. Phys.}\ }\textbf {\bibinfo
  {volume} {14}},\ \bibinfo {pages} {125003} (\bibinfo {year} {2012})},\
  \Eprint {http://arxiv.org/abs/1206.2942} {arXiv:1206.2942 [hep-ph]}
  \BibitemShut {NoStop}%
\bibitem [{\citenamefont {Brax}\ \emph {et~al.}(2004)\citenamefont {Brax},
  \citenamefont {van~de Bruck}, \citenamefont {Davis}, \citenamefont {Khoury},\
  and\ \citenamefont {Weltman}}]{Brax:2004qh}%
  \BibitemOpen
  \bibfield  {author} {\bibinfo {author} {\bibfnamefont {P.}~\bibnamefont
  {Brax}}, \bibinfo {author} {\bibfnamefont {C.}~\bibnamefont {van~de Bruck}},
  \bibinfo {author} {\bibfnamefont {A.-C.}\ \bibnamefont {Davis}}, \bibinfo
  {author} {\bibfnamefont {J.}~\bibnamefont {Khoury}}, \ and\ \bibinfo {author}
  {\bibfnamefont {A.}~\bibnamefont {Weltman}},\ }\href {\doibase
  10.1103/PhysRevD.70.123518} {\bibfield  {journal} {\bibinfo  {journal} {Phys.
  Rev.}\ }\textbf {\bibinfo {volume} {D70}},\ \bibinfo {pages} {123518}
  (\bibinfo {year} {2004})},\ \Eprint {http://arxiv.org/abs/astro-ph/0408415}
  {arXiv:astro-ph/0408415 [astro-ph]} \BibitemShut {NoStop}%
\bibitem [{\citenamefont {Erickcek}\ \emph {et~al.}(2013)\citenamefont
  {Erickcek}, \citenamefont {Barnaby}, \citenamefont {Burrage},\ and\
  \citenamefont {Huang}}]{Erickcek:2013oma}%
  \BibitemOpen
  \bibfield  {author} {\bibinfo {author} {\bibfnamefont {A.~L.}\ \bibnamefont
  {Erickcek}}, \bibinfo {author} {\bibfnamefont {N.}~\bibnamefont {Barnaby}},
  \bibinfo {author} {\bibfnamefont {C.}~\bibnamefont {Burrage}}, \ and\
  \bibinfo {author} {\bibfnamefont {Z.}~\bibnamefont {Huang}},\ }\href
  {\doibase 10.1103/PhysRevLett.110.171101} {\bibfield  {journal} {\bibinfo
  {journal} {Phys. Rev. Lett.}\ }\textbf {\bibinfo {volume} {110}},\ \bibinfo
  {pages} {171101} (\bibinfo {year} {2013})},\ \Eprint
  {http://arxiv.org/abs/1304.0009} {arXiv:1304.0009 [astro-ph.CO]} \BibitemShut
  {NoStop}%
\bibitem [{\citenamefont {Capozziello}\ \emph {et~al.}(2018)\citenamefont
  {Capozziello}, \citenamefont {Nojiri},\ and\ \citenamefont
  {Odintsov}}]{Capozziello:2018wul}%
  \BibitemOpen
  \bibfield  {author} {\bibinfo {author} {\bibfnamefont {S.}~\bibnamefont
  {Capozziello}}, \bibinfo {author} {\bibfnamefont {S.}~\bibnamefont {Nojiri}},
  \ and\ \bibinfo {author} {\bibfnamefont {S.~D.}\ \bibnamefont {Odintsov}},\
  }\href {\doibase 10.1016/j.physletb.2018.03.064} {\bibfield  {journal}
  {\bibinfo  {journal} {Phys. Lett.}\ }\textbf {\bibinfo {volume} {B781}},\
  \bibinfo {pages} {99} (\bibinfo {year} {2018})},\ \Eprint
  {http://arxiv.org/abs/1803.08815} {arXiv:1803.08815 [gr-qc]} \BibitemShut
  {NoStop}%
\bibitem [{\citenamefont {Gildener}\ and\ \citenamefont
  {Weinberg}(1976)}]{Gildener:1976ih}%
  \BibitemOpen
  \bibfield  {author} {\bibinfo {author} {\bibfnamefont {E.}~\bibnamefont
  {Gildener}}\ and\ \bibinfo {author} {\bibfnamefont {S.}~\bibnamefont
  {Weinberg}},\ }\href {\doibase 10.1103/PhysRevD.13.3333} {\bibfield
  {journal} {\bibinfo  {journal} {Phys. Rev.}\ }\textbf {\bibinfo {volume}
  {D13}},\ \bibinfo {pages} {3333} (\bibinfo {year} {1976})}\BibitemShut
  {NoStop}%
\bibitem [{\citenamefont {Branco}\ \emph {et~al.}(2012)\citenamefont {Branco},
  \citenamefont {Ferreira}, \citenamefont {Lavoura}, \citenamefont {Rebelo},
  \citenamefont {Sher},\ and\ \citenamefont {Silva}}]{Branco:2011iw}%
  \BibitemOpen
  \bibfield  {author} {\bibinfo {author} {\bibfnamefont {G.~C.}\ \bibnamefont
  {Branco}}, \bibinfo {author} {\bibfnamefont {P.~M.}\ \bibnamefont
  {Ferreira}}, \bibinfo {author} {\bibfnamefont {L.}~\bibnamefont {Lavoura}},
  \bibinfo {author} {\bibfnamefont {M.~N.}\ \bibnamefont {Rebelo}}, \bibinfo
  {author} {\bibfnamefont {M.}~\bibnamefont {Sher}}, \ and\ \bibinfo {author}
  {\bibfnamefont {J.~P.}\ \bibnamefont {Silva}},\ }\href {\doibase
  10.1016/j.physrep.2012.02.002} {\bibfield  {journal} {\bibinfo  {journal}
  {Phys. Rept.}\ }\textbf {\bibinfo {volume} {516}},\ \bibinfo {pages} {1}
  (\bibinfo {year} {2012})},\ \Eprint {http://arxiv.org/abs/1106.0034}
  {arXiv:1106.0034 [hep-ph]} \BibitemShut {NoStop}%
\bibitem [{\citenamefont {Chiang}\ and\ \citenamefont
  {Senaha}(2017)}]{Chiang:2017zbz}%
  \BibitemOpen
  \bibfield  {author} {\bibinfo {author} {\bibfnamefont {C.-W.}\ \bibnamefont
  {Chiang}}\ and\ \bibinfo {author} {\bibfnamefont {E.}~\bibnamefont
  {Senaha}},\ }\href {\doibase 10.1016/j.physletb.2017.09.064} {\bibfield
  {journal} {\bibinfo  {journal} {Phys. Lett.}\ }\textbf {\bibinfo {volume}
  {B774}},\ \bibinfo {pages} {489} (\bibinfo {year} {2017})},\ \Eprint
  {http://arxiv.org/abs/1707.06765} {arXiv:1707.06765 [hep-ph]} \BibitemShut
  {NoStop}%
\bibitem [{\citenamefont {Patel}\ and\ \citenamefont
  {Ramsey-Musolf}(2011)}]{Patel:2011th}%
  \BibitemOpen
  \bibfield  {author} {\bibinfo {author} {\bibfnamefont {H.~H.}\ \bibnamefont
  {Patel}}\ and\ \bibinfo {author} {\bibfnamefont {M.~J.}\ \bibnamefont
  {Ramsey-Musolf}},\ }\href {\doibase 10.1007/JHEP07(2011)029} {\bibfield
  {journal} {\bibinfo  {journal} {JHEP}\ }\textbf {\bibinfo {volume} {07}},\
  \bibinfo {pages} {029} (\bibinfo {year} {2011})},\ \Eprint
  {http://arxiv.org/abs/1101.4665} {arXiv:1101.4665 [hep-ph]} \BibitemShut
  {NoStop}%
\bibitem [{\citenamefont {Parwani}(1992)}]{Parwani:1991gq}%
  \BibitemOpen
  \bibfield  {author} {\bibinfo {author} {\bibfnamefont {R.~R.}\ \bibnamefont
  {Parwani}},\ }\href {\doibase 10.1103/PhysRevD.45.4695,
  10.1103/PhysRevD.48.5965.2} {\bibfield  {journal} {\bibinfo  {journal} {Phys.
  Rev.}\ }\textbf {\bibinfo {volume} {D45}},\ \bibinfo {pages} {4695} (\bibinfo
  {year} {1992})},\ \bibinfo {note} {[Erratum: Phys. Rev.D48,5965(1993)]},\
  \Eprint {http://arxiv.org/abs/hep-ph/9204216} {arXiv:hep-ph/9204216 [hep-ph]}
  \BibitemShut {NoStop}%
\bibitem [{\citenamefont {Buchmuller}\ \emph {et~al.}(1993)\citenamefont
  {Buchmuller}, \citenamefont {Helbig},\ and\ \citenamefont
  {Walliser}}]{Buchmuller:1992rs}%
  \BibitemOpen
  \bibfield  {author} {\bibinfo {author} {\bibfnamefont {W.}~\bibnamefont
  {Buchmuller}}, \bibinfo {author} {\bibfnamefont {T.}~\bibnamefont {Helbig}},
  \ and\ \bibinfo {author} {\bibfnamefont {D.}~\bibnamefont {Walliser}},\
  }\href {\doibase 10.1016/0550-3213(93)90064-V} {\bibfield  {journal}
  {\bibinfo  {journal} {Nucl. Phys.}\ }\textbf {\bibinfo {volume} {B407}},\
  \bibinfo {pages} {387} (\bibinfo {year} {1993})}\BibitemShut {NoStop}%
\bibitem [{\citenamefont {Chiku}\ and\ \citenamefont
  {Hatsuda}(1998)}]{Chiku:1998kd}%
  \BibitemOpen
  \bibfield  {author} {\bibinfo {author} {\bibfnamefont {S.}~\bibnamefont
  {Chiku}}\ and\ \bibinfo {author} {\bibfnamefont {T.}~\bibnamefont
  {Hatsuda}},\ }\href {\doibase 10.1103/PhysRevD.58.076001} {\bibfield
  {journal} {\bibinfo  {journal} {Phys. Rev.}\ }\textbf {\bibinfo {volume}
  {D58}},\ \bibinfo {pages} {076001} (\bibinfo {year} {1998})},\ \Eprint
  {http://arxiv.org/abs/hep-ph/9803226} {arXiv:hep-ph/9803226 [hep-ph]}
  \BibitemShut {NoStop}%
\bibitem [{\citenamefont {Funakubo}\ and\ \citenamefont
  {Senaha}(2013)}]{Funakubo:2012qc}%
  \BibitemOpen
  \bibfield  {author} {\bibinfo {author} {\bibfnamefont {K.}~\bibnamefont
  {Funakubo}}\ and\ \bibinfo {author} {\bibfnamefont {E.}~\bibnamefont
  {Senaha}},\ }\href {\doibase 10.1103/PhysRevD.87.054003} {\bibfield
  {journal} {\bibinfo  {journal} {Phys. Rev.}\ }\textbf {\bibinfo {volume}
  {D87}},\ \bibinfo {pages} {054003} (\bibinfo {year} {2013})},\ \Eprint
  {http://arxiv.org/abs/1210.1737} {arXiv:1210.1737 [hep-ph]} \BibitemShut
  {NoStop}%
\bibitem [{\citenamefont {Carrington}(1992)}]{Carrington:1991hz}%
  \BibitemOpen
  \bibfield  {author} {\bibinfo {author} {\bibfnamefont {M.~E.}\ \bibnamefont
  {Carrington}},\ }\href {\doibase 10.1103/PhysRevD.45.2933} {\bibfield
  {journal} {\bibinfo  {journal} {Phys. Rev.}\ }\textbf {\bibinfo {volume}
  {D45}},\ \bibinfo {pages} {2933} (\bibinfo {year} {1992})}\BibitemShut
  {NoStop}%
\bibitem [{\citenamefont {Funakubo}\ and\ \citenamefont
  {Senaha}(2009)}]{Funakubo:2009eg}%
  \BibitemOpen
  \bibfield  {author} {\bibinfo {author} {\bibfnamefont {K.}~\bibnamefont
  {Funakubo}}\ and\ \bibinfo {author} {\bibfnamefont {E.}~\bibnamefont
  {Senaha}},\ }\href {\doibase 10.1103/PhysRevD.79.115024} {\bibfield
  {journal} {\bibinfo  {journal} {Phys. Rev.}\ }\textbf {\bibinfo {volume}
  {D79}},\ \bibinfo {pages} {115024} (\bibinfo {year} {2009})},\ \Eprint
  {http://arxiv.org/abs/0905.2022} {arXiv:0905.2022 [hep-ph]} \BibitemShut
  {NoStop}%
\bibitem [{\citenamefont {Manton}(1983)}]{Manton:1983nd}%
  \BibitemOpen
  \bibfield  {author} {\bibinfo {author} {\bibfnamefont {N.~S.}\ \bibnamefont
  {Manton}},\ }\href {\doibase 10.1103/PhysRevD.28.2019} {\bibfield  {journal}
  {\bibinfo  {journal} {Phys. Rev.}\ }\textbf {\bibinfo {volume} {D28}},\
  \bibinfo {pages} {2019} (\bibinfo {year} {1983})}\BibitemShut {NoStop}%
\bibitem [{\citenamefont {Klinkhamer}\ and\ \citenamefont
  {Manton}(1984)}]{Klinkhamer:1984di}%
  \BibitemOpen
  \bibfield  {author} {\bibinfo {author} {\bibfnamefont {F.~R.}\ \bibnamefont
  {Klinkhamer}}\ and\ \bibinfo {author} {\bibfnamefont {N.~S.}\ \bibnamefont
  {Manton}},\ }\href {\doibase 10.1103/PhysRevD.30.2212} {\bibfield  {journal}
  {\bibinfo  {journal} {Phys. Rev.}\ }\textbf {\bibinfo {volume} {D30}},\
  \bibinfo {pages} {2212} (\bibinfo {year} {1984})}\BibitemShut {NoStop}%
\bibitem [{\citenamefont {Linde}(1983)}]{Linde:1981zj}%
  \BibitemOpen
  \bibfield  {author} {\bibinfo {author} {\bibfnamefont {A.~D.}\ \bibnamefont
  {Linde}},\ }\href {\doibase 10.1016/0550-3213(83)90293-6,
  10.1016/0550-3213(83)90072-X} {\bibfield  {journal} {\bibinfo  {journal}
  {Nucl. Phys.}\ }\textbf {\bibinfo {volume} {B216}},\ \bibinfo {pages} {421}
  (\bibinfo {year} {1983})},\ \bibinfo {note} {[Erratum: Nucl.
  Phys.B223,544(1983)]}\BibitemShut {NoStop}%
\bibitem [{\citenamefont {Andreev}\ \emph {et~al.}(2018)\citenamefont {Andreev}
  \emph {et~al.}}]{Andreev:2018ayy}%
  \BibitemOpen
  \bibfield  {author} {\bibinfo {author} {\bibfnamefont {V.}~\bibnamefont
  {Andreev}} \emph {et~al.} (\bibinfo {collaboration} {ACME}),\ }\href
  {\doibase 10.1038/s41586-018-0599-8} {\bibfield  {journal} {\bibinfo
  {journal} {Nature}\ }\textbf {\bibinfo {volume} {562}},\ \bibinfo {pages}
  {355} (\bibinfo {year} {2018})}\BibitemShut {NoStop}%
\bibitem [{\citenamefont {Katsuragawa}\ and\ \citenamefont
  {Matsuzaki}(2017)}]{Katsuragawa:2016yir}%
  \BibitemOpen
  \bibfield  {author} {\bibinfo {author} {\bibfnamefont {T.}~\bibnamefont
  {Katsuragawa}}\ and\ \bibinfo {author} {\bibfnamefont {S.}~\bibnamefont
  {Matsuzaki}},\ }\href {\doibase 10.1103/PhysRevD.95.044040} {\bibfield
  {journal} {\bibinfo  {journal} {Phys. Rev.}\ }\textbf {\bibinfo {volume}
  {D95}},\ \bibinfo {pages} {044040} (\bibinfo {year} {2017})},\ \Eprint
  {http://arxiv.org/abs/1610.01016} {arXiv:1610.01016 [gr-qc]} \BibitemShut
  {NoStop}%
\bibitem [{\citenamefont {Burrage}\ \emph {et~al.}(2018)\citenamefont
  {Burrage}, \citenamefont {Copeland}, \citenamefont {Millington},\ and\
  \citenamefont {Spannowsky}}]{Burrage:2018dvt}%
  \BibitemOpen
  \bibfield  {author} {\bibinfo {author} {\bibfnamefont {C.}~\bibnamefont
  {Burrage}}, \bibinfo {author} {\bibfnamefont {E.~J.}\ \bibnamefont
  {Copeland}}, \bibinfo {author} {\bibfnamefont {P.}~\bibnamefont
  {Millington}}, \ and\ \bibinfo {author} {\bibfnamefont {M.}~\bibnamefont
  {Spannowsky}},\ }\href {\doibase 10.1088/1475-7516/2018/11/036} {\bibfield
  {journal} {\bibinfo  {journal} {JCAP}\ }\textbf {\bibinfo {volume} {1811}},\
  \bibinfo {pages} {036} (\bibinfo {year} {2018})},\ \Eprint
  {http://arxiv.org/abs/1804.07180} {arXiv:1804.07180 [hep-th]} \BibitemShut
  {NoStop}%
\bibitem [{\citenamefont {Starobinsky}(2007)}]{Starobinsky:2007hu}%
  \BibitemOpen
  \bibfield  {author} {\bibinfo {author} {\bibfnamefont {A.~A.}\ \bibnamefont
  {Starobinsky}},\ }\href {\doibase 10.1134/S0021364007150027} {\bibfield
  {journal} {\bibinfo  {journal} {JETP Lett.}\ }\textbf {\bibinfo {volume}
  {86}},\ \bibinfo {pages} {157} (\bibinfo {year} {2007})},\ \Eprint
  {http://arxiv.org/abs/0706.2041} {arXiv:0706.2041 [astro-ph]} \BibitemShut
  {NoStop}%
\bibitem [{\citenamefont {Frolov}(2008)}]{Frolov:2008uf}%
  \BibitemOpen
  \bibfield  {author} {\bibinfo {author} {\bibfnamefont {A.~V.}\ \bibnamefont
  {Frolov}},\ }\href {\doibase 10.1103/PhysRevLett.101.061103} {\bibfield
  {journal} {\bibinfo  {journal} {Phys. Rev. Lett.}\ }\textbf {\bibinfo
  {volume} {101}},\ \bibinfo {pages} {061103} (\bibinfo {year} {2008})},\
  \Eprint {http://arxiv.org/abs/0803.2500} {arXiv:0803.2500 [astro-ph]}
  \BibitemShut {NoStop}%
\bibitem [{\citenamefont {Dev}\ \emph {et~al.}(2008)\citenamefont {Dev},
  \citenamefont {Jain}, \citenamefont {Jhingan}, \citenamefont {Nojiri},
  \citenamefont {Sami},\ and\ \citenamefont {Thongkool}}]{Dev:2008rx}%
  \BibitemOpen
  \bibfield  {author} {\bibinfo {author} {\bibfnamefont {A.}~\bibnamefont
  {Dev}}, \bibinfo {author} {\bibfnamefont {D.}~\bibnamefont {Jain}}, \bibinfo
  {author} {\bibfnamefont {S.}~\bibnamefont {Jhingan}}, \bibinfo {author}
  {\bibfnamefont {S.}~\bibnamefont {Nojiri}}, \bibinfo {author} {\bibfnamefont
  {M.}~\bibnamefont {Sami}}, \ and\ \bibinfo {author} {\bibfnamefont
  {I.}~\bibnamefont {Thongkool}},\ }\href {\doibase 10.1103/PhysRevD.78.083515}
  {\bibfield  {journal} {\bibinfo  {journal} {Phys. Rev.}\ }\textbf {\bibinfo
  {volume} {D78}},\ \bibinfo {pages} {083515} (\bibinfo {year} {2008})},\
  \Eprint {http://arxiv.org/abs/0807.3445} {arXiv:0807.3445 [hep-th]}
  \BibitemShut {NoStop}%
\bibitem [{\citenamefont {Kobayashi}\ and\ \citenamefont
  {Maeda}(2009)}]{Kobayashi:2008wc}%
  \BibitemOpen
  \bibfield  {author} {\bibinfo {author} {\bibfnamefont {T.}~\bibnamefont
  {Kobayashi}}\ and\ \bibinfo {author} {\bibfnamefont {K.-i.}\ \bibnamefont
  {Maeda}},\ }\href {\doibase 10.1103/PhysRevD.79.024009} {\bibfield  {journal}
  {\bibinfo  {journal} {Phys. Rev.}\ }\textbf {\bibinfo {volume} {D79}},\
  \bibinfo {pages} {024009} (\bibinfo {year} {2009})},\ \Eprint
  {http://arxiv.org/abs/0810.5664} {arXiv:0810.5664 [astro-ph]} \BibitemShut
  {NoStop}%
\bibitem [{\citenamefont {Thongkool}\ \emph {et~al.}(2009)\citenamefont
  {Thongkool}, \citenamefont {Sami}, \citenamefont {Gannouji},\ and\
  \citenamefont {Jhingan}}]{Thongkool:2009js}%
  \BibitemOpen
  \bibfield  {author} {\bibinfo {author} {\bibfnamefont {I.}~\bibnamefont
  {Thongkool}}, \bibinfo {author} {\bibfnamefont {M.}~\bibnamefont {Sami}},
  \bibinfo {author} {\bibfnamefont {R.}~\bibnamefont {Gannouji}}, \ and\
  \bibinfo {author} {\bibfnamefont {S.}~\bibnamefont {Jhingan}},\ }\href
  {\doibase 10.1103/PhysRevD.80.043523} {\bibfield  {journal} {\bibinfo
  {journal} {Phys. Rev.}\ }\textbf {\bibinfo {volume} {D80}},\ \bibinfo {pages}
  {043523} (\bibinfo {year} {2009})},\ \Eprint {http://arxiv.org/abs/0906.2460}
  {arXiv:0906.2460 [hep-th]} \BibitemShut {NoStop}%
\bibitem [{\citenamefont {Cembranos}(2009)}]{Cembranos:2008gj}%
  \BibitemOpen
  \bibfield  {author} {\bibinfo {author} {\bibfnamefont {J.~A.~R.}\
  \bibnamefont {Cembranos}},\ }\href {\doibase 10.1103/PhysRevLett.102.141301}
  {\bibfield  {journal} {\bibinfo  {journal} {Phys. Rev. Lett.}\ }\textbf
  {\bibinfo {volume} {102}},\ \bibinfo {pages} {141301} (\bibinfo {year}
  {2009})},\ \Eprint {http://arxiv.org/abs/0809.1653} {arXiv:0809.1653
  [hep-ph]} \BibitemShut {NoStop}%
\bibitem [{\citenamefont {Corda}(2008)}]{Corda:2008jt}%
  \BibitemOpen
  \bibfield  {author} {\bibinfo {author} {\bibfnamefont {C.}~\bibnamefont
  {Corda}},\ }\href {\doibase 10.1007/s10714-008-0627-3} {\bibfield  {journal}
  {\bibinfo  {journal} {Gen. Rel. Grav.}\ }\textbf {\bibinfo {volume} {40}},\
  \bibinfo {pages} {2201} (\bibinfo {year} {2008})},\ \Eprint
  {http://arxiv.org/abs/0802.2523} {arXiv:0802.2523 [astro-ph]} \BibitemShut
  {NoStop}%
\bibitem [{\citenamefont {Capozziello}\ \emph {et~al.}(2008)\citenamefont
  {Capozziello}, \citenamefont {Corda},\ and\ \citenamefont
  {De~Laurentis}}]{Capozziello:2008rq}%
  \BibitemOpen
  \bibfield  {author} {\bibinfo {author} {\bibfnamefont {S.}~\bibnamefont
  {Capozziello}}, \bibinfo {author} {\bibfnamefont {C.}~\bibnamefont {Corda}},
  \ and\ \bibinfo {author} {\bibfnamefont {M.~F.}\ \bibnamefont
  {De~Laurentis}},\ }\href {\doibase 10.1016/j.physletb.2008.10.001} {\bibfield
   {journal} {\bibinfo  {journal} {Phys. Lett.}\ }\textbf {\bibinfo {volume}
  {B669}},\ \bibinfo {pages} {255} (\bibinfo {year} {2008})},\ \Eprint
  {http://arxiv.org/abs/0812.2272} {arXiv:0812.2272 [astro-ph]} \BibitemShut
  {NoStop}%
\bibitem [{\citenamefont {Kosowsky}\ \emph
  {et~al.}(1992{\natexlab{a}})\citenamefont {Kosowsky}, \citenamefont
  {Turner},\ and\ \citenamefont {Watkins}}]{Kosowsky:1991ua}%
  \BibitemOpen
  \bibfield  {author} {\bibinfo {author} {\bibfnamefont {A.}~\bibnamefont
  {Kosowsky}}, \bibinfo {author} {\bibfnamefont {M.~S.}\ \bibnamefont
  {Turner}}, \ and\ \bibinfo {author} {\bibfnamefont {R.}~\bibnamefont
  {Watkins}},\ }\href {\doibase 10.1103/PhysRevD.45.4514} {\bibfield  {journal}
  {\bibinfo  {journal} {Phys. Rev.}\ }\textbf {\bibinfo {volume} {D45}},\
  \bibinfo {pages} {4514} (\bibinfo {year} {1992}{\natexlab{a}})}\BibitemShut
  {NoStop}%
\bibitem [{\citenamefont {Kosowsky}\ \emph
  {et~al.}(1992{\natexlab{b}})\citenamefont {Kosowsky}, \citenamefont
  {Turner},\ and\ \citenamefont {Watkins}}]{Kosowsky:1992rz}%
  \BibitemOpen
  \bibfield  {author} {\bibinfo {author} {\bibfnamefont {A.}~\bibnamefont
  {Kosowsky}}, \bibinfo {author} {\bibfnamefont {M.~S.}\ \bibnamefont
  {Turner}}, \ and\ \bibinfo {author} {\bibfnamefont {R.}~\bibnamefont
  {Watkins}},\ }\href {\doibase 10.1103/PhysRevLett.69.2026} {\bibfield
  {journal} {\bibinfo  {journal} {Phys. Rev. Lett.}\ }\textbf {\bibinfo
  {volume} {69}},\ \bibinfo {pages} {2026} (\bibinfo {year}
  {1992}{\natexlab{b}})}\BibitemShut {NoStop}%
\bibitem [{\citenamefont {Kosowsky}\ and\ \citenamefont
  {Turner}(1993)}]{Kosowsky:1992vn}%
  \BibitemOpen
  \bibfield  {author} {\bibinfo {author} {\bibfnamefont {A.}~\bibnamefont
  {Kosowsky}}\ and\ \bibinfo {author} {\bibfnamefont {M.~S.}\ \bibnamefont
  {Turner}},\ }\href {\doibase 10.1103/PhysRevD.47.4372} {\bibfield  {journal}
  {\bibinfo  {journal} {Phys. Rev.}\ }\textbf {\bibinfo {volume} {D47}},\
  \bibinfo {pages} {4372} (\bibinfo {year} {1993})},\ \Eprint
  {http://arxiv.org/abs/astro-ph/9211004} {arXiv:astro-ph/9211004 [astro-ph]}
  \BibitemShut {NoStop}%
\bibitem [{\citenamefont {Kamionkowski}\ \emph {et~al.}(1994)\citenamefont
  {Kamionkowski}, \citenamefont {Kosowsky},\ and\ \citenamefont
  {Turner}}]{Kamionkowski:1993fg}%
  \BibitemOpen
  \bibfield  {author} {\bibinfo {author} {\bibfnamefont {M.}~\bibnamefont
  {Kamionkowski}}, \bibinfo {author} {\bibfnamefont {A.}~\bibnamefont
  {Kosowsky}}, \ and\ \bibinfo {author} {\bibfnamefont {M.~S.}\ \bibnamefont
  {Turner}},\ }\href {\doibase 10.1103/PhysRevD.49.2837} {\bibfield  {journal}
  {\bibinfo  {journal} {Phys. Rev.}\ }\textbf {\bibinfo {volume} {D49}},\
  \bibinfo {pages} {2837} (\bibinfo {year} {1994})},\ \Eprint
  {http://arxiv.org/abs/astro-ph/9310044} {arXiv:astro-ph/9310044 [astro-ph]}
  \BibitemShut {NoStop}%
\end{thebibliography}%

\end{document}